\documentclass[12pt]{JHEP3}
\usepackage{epsfig}
\usepackage{cite}
\newcommand{\nparton}{n}
\title{\bf  Complete off-shell effects in top quark pair hadroproduction 
with leptonic decay at next-to-leading order}

\author{
Giuseppe Bevilacqua$^{a,b}$, Micha\l{}  Czakon$^{b}$, 
Andreas van Hameren$^{c}$, Costas G. Papadopoulos$^{a}$ 
and Ma\l{}gorzata Worek$^{d}$ \\ \\

$^a$~{Institute of Nuclear Physics, NCSR Demokritos,
      GR-15310 Athens, Greece}\\
$^b$~{
Institut f\"ur Theoretische Teilchenphysik und Kosmologie, 
      RWTH Aachen University,
      D-52056 Aachen, Germany}\\
$^c$~{The H. Niewodnicza\'nski Institute of Nuclear Physics,
      Polisch Academy of Sciences,
      Radzikowskiego 152, PL-31342 Cracow, Poland}\\
$^d$~{Fachbereich C Physik, Bergische Universit\"at Wuppertal, D-42097 
      Wuppertal,Germany}}

\abstract{  Results for next-to-leading order QCD corrections to the
  $pp(p\bar{p}) \to t \bar{t}\to W^+W^- b\bar{b} \to e^{+}\nu_{e}
  \mu^{-}\bar{\nu}_{\mu}b\bar{b} ~+X$  processes with complete
  off-shell effects are presented for the first time.  Double-,
  single- and non-resonant top contributions of  the order
  ${\cal{O}}(\alpha_{s}^3 \alpha^4)$  are consistently taken into
  account, which requires the introduction of a complex-mass scheme
  for unstable  top quarks.  Moreover, the intermediate $W$ bosons are
  treated off-shell.   Comparison to the narrow width approximation
  for top quarks, where non-factorizable corrections are not accounted
  for is performed.   Besides the total cross section and its scale
  dependence, several differential distributions at the TeVatron run
  II and the LHC are given.    In case of the TeVatron the
  forward-backward asymmetry of the top is recalculated afresh.  With
  inclusive selection cuts,  the forward-backward asymmetry   amounts
  to $A^{t}_{FB} = 0.051 \pm 0.0013$. Furthermore, the corrections
  with respect to leading order are positive  and of the order $2.3
  \%$  for the TeVatron and $47\%$ for the LHC.  A study of the scale
  dependence of our NLO predictions indicates that the residual
  theoretical uncertainty due to higher order corrections is $8\%$ for
  the TeVatron and $9\%$ for the LHC.  }

\preprint{TTK-10-56\\
          IFJPAN-IV-2010-9 \\ 
          WUB/10-26}

\begin{document}

\section{Introduction}
\label{sec:introduction}

The $t\bar{t}$ production process is a copious source of $W$-pairs
and, hence, of  isolated leptons at the TeVatron and the LHC. In
consequence it is intensely studied  as a signal at these
colliders. In view of the   large production rate, precise and direct
measurements are possible,  which require a  detailed theoretical
understanding.  In addition, it constitutes an important  background
for many new particle searches. Examples include the leptonic  signals
for cascade decays of supersymmetric particles or searches for  $H\to
W^+W^−$ and $H\to \tau^+\tau^-$ decays.    

Even though, the first results for next-to-leading order (NLO) QCD
corrections to heavy quark production were presented more than twenty
years ago
\cite{Nason:1987xz,Beenakker:1988bq,Nason:1989zy,Beenakker:1990maa},
recent progress in NLO  \cite{Bernreuther:2001rq,Bernreuther:2004jv,
  Bernreuther:2010ny,Melnikov:2009dn,Czakon:2008ii}  and
next-to-next-to leading order (NNLO)
\cite{Czakon:2008zk,Bonciani:2008az,Anastasiou:2008vd,Kniehl:2008fd,
  Bonciani:2009nb,Bonciani:2010mn} calculations, as well as
next-to-next-to-leading-log resummations (NNLL)
\cite{Czakon:2008cx,Kidonakis:2009ev,Beneke:2009rj,Czakon:2009zw,
Beneke:2009ye,Ahrens:2010zv,Kidonakis:2010dk}  
for inclusive $t\bar{t}$ hadroproduction is truly astonishing.

The list for the  more exclusive channels is just as impressive: NLO
QCD corrections have been calculated for the $t\bar{t}H$ signal
\cite{Beenakker:2001rj,Reina:2001bc,Reina:2001sf,Beenakker:2002nc,
  Dawson:2002tg,Dawson:2003zu}, where the Higgs boson has been treated
as a stable particle. Most recently  the factorizable QCD corrections
to this process have been presented \cite{Binoth:2010ra}, where higher
order corrections  to both production and decay of the Higgs boson to
a $b\bar{b} $ pair have been calculated with the latter modeled  by
the Higgs propagator with a fixed width.  Moreover, NLO QCD
corrections to a variety of $2 \to 3$ backgrounds processes
$t\bar{t}j$ \cite{Dittmaier:2007wz,Dittmaier:2008uj, Melnikov:2010iu},
$t\bar{t}Z$ \cite{Lazopoulos:2008de} and  $t\bar{t}\gamma$
\cite{PengFei:2009ph} have been obtained. Most recently, NLO QCD
corrections to    $2 \to 4$ backgrounds $t\bar{t}b\bar{b}$
\cite{Bredenstein:2008zb,Bredenstein:2009aj,
  Bevilacqua:2009zn,Bredenstein:2010rs}  and $t\bar{t}jj$
\cite{Bevilacqua:2010ve} have also been evaluated.

Usually, $t\bar{t}$ production is restricted to on-shell states and
decays if available are treated   in the narrow-width approximation
(NWA), which effectively decouples top production and decay.  The NWA
allows to neglect non-resonant as well as non-factorizable amplitude
contributions,  thus leading to significant simplifications for
calculations of higher order corrections.  Whenever resonant  top
production dominates, as it does for very inclusive cuts,
this approximation is of course well motivated.  In some cases
calculations have been further simplified by also treating the
decaying $W$ bosons as on-shell particles.

Naturally, the accuracy of these approximations needs to be tested,
which  requires a full calculation of off-shell effects.  One thus
needs a calculation which includes both resonant and non-resonant
contributions, using finite width top-quark propagators,  which
correctly includes interference effects between the various
contributions. The purpose of this paper is to present such a complete
calculation for $t\bar{t}$ production at NLO QCD level.  In addition
to merging resonant and non-resonant effects for the top quarks,  we
also include finite width effects for the $W$ bosons, {\it i.e.} we
consider NLO QCD corrections to the  general
$e^+\nu_e\mu^-\bar{\nu}_{\mu}b\bar{b}$ final state.

In addition, all selection strategies based on next-to-leading order
simulations, which have been devised for the efficient suppression of
$t\bar{t}$ background, are at present optimized against top production
in the NWA. Within our approach, presented in
the form of a flexible  Monte Carlo  program which allows to study NLO
QCD corrections to cross sections and kinematic distributions with
arbitrary cuts   on particles in the final state and  with full spin
correlations,  it is possible to reexamine the quality of the chosen
selection with improved accuracy.

The paper is organized as follows. In Section  \ref{sec:theoretical
  framework} we briefly describe the calculation of the NLO
corrections. Numerical results for the integrated and differential
cross sections  are presented in Section \ref{sec:results} both for
the TeVatron and the LHC. Finally, we conclude in Section
\ref{sec:conclusions}.

\section{Theoretical framework}

\label{sec:theoretical framework}

\subsection{Born level}

\begin{figure}
\begin{center}
\includegraphics[width=0.99\textwidth]{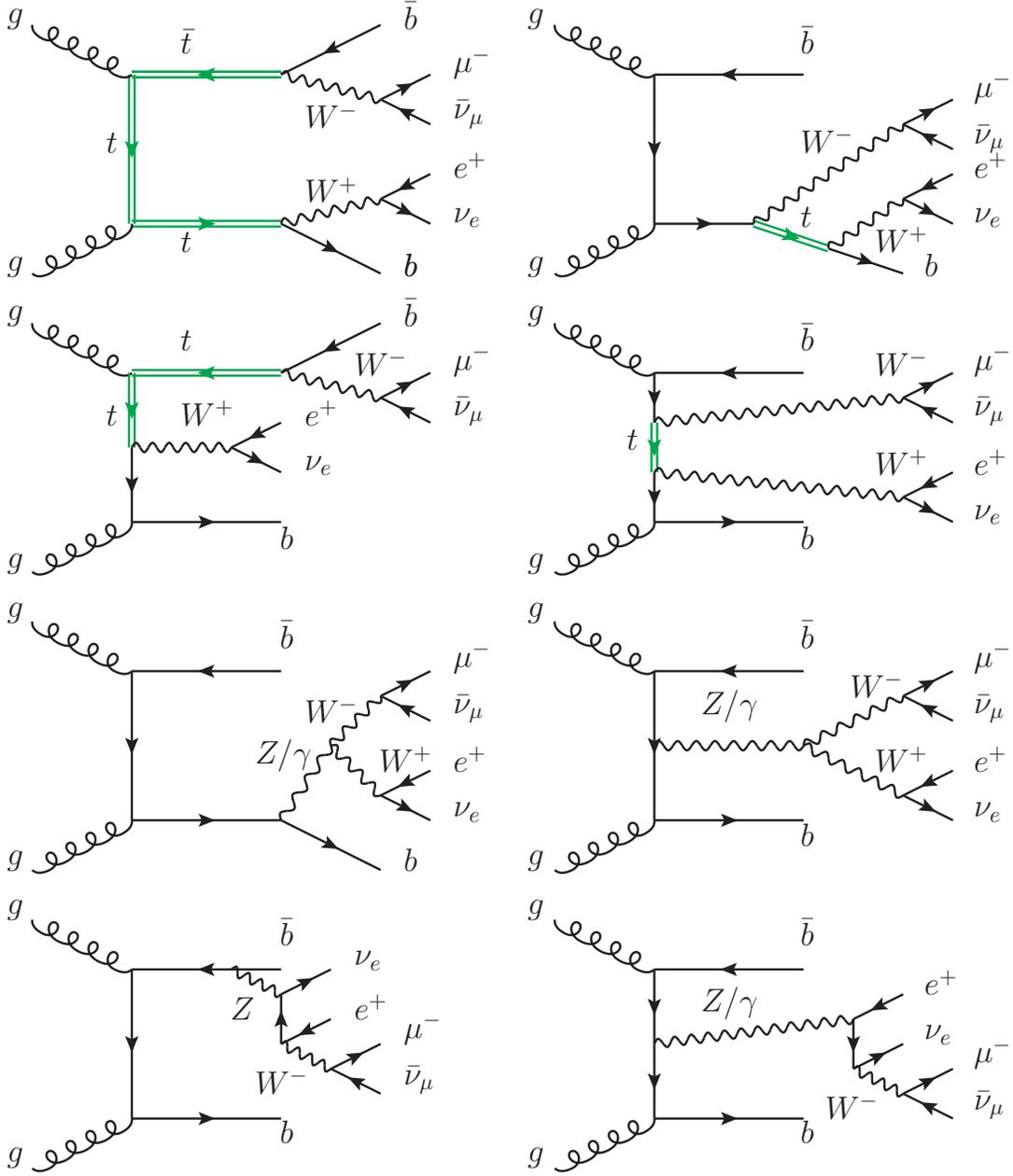}
\end{center}
\caption{\it \label{fig:diagrams-lo}   Representative Feynman diagrams
  contributing to the leading order process $gg\to
  e^{+}\nu_{e}\mu^{-}\bar{\nu}_{\mu}b\bar{b} $  at
  ${\cal{O}}(\alpha_{s}^2 \alpha^4)$,  with different off-shell
  intermediate states:  double-, single-, and non-resonant
  top quark contributions.}
\end{figure}

At Born level the partonic reactions are   
\[
gg\to e^{+}\nu_{e}\mu^{-}\bar{\nu}_{\mu}b\bar{b}
\] 
\begin{equation}
q\bar{q}\to e^{+}\nu_{e}\mu^{-}\bar{\nu}_{\mu}b\bar{b} 
\end{equation}
where $q$ stands for up- or down-type quarks. The
${\cal{O}}(\alpha_{s}^2 \alpha^4)$ contributions to the
$e^{+}\nu_{e}\mu^{-}\bar{\nu}_{\mu}b\bar{b}$  process  can be
subdivided into three classes, namely diagrams containing two top
quark propagators that can become resonant,  diagrams containing only
one top quark resonance  and finally diagrams without any top quark
resonance. Regarding the $W^{\pm}$ resonances one can distinguish only
two subclasses, double- and single-resonant gauge boson
contributions. A few examples of Feynman diagrams  contributing to the
leading order $gg\to e^{+}\nu_{e}\mu^{-}\bar{\nu}_{\mu}b\bar{b}$
subprocess are presented in Figure \ref{fig:diagrams-lo}.  

Since the produced top quarks are unstable particles, the inclusion of
the decays is performed in the complex mass scheme, which for LO is
described in Ref. \cite{Denner:1999gp,Denner:2005fg}.  It fully
respects gauge invariance and is straightforward to apply. In the
amplitude (at LO and NLO) we simply perform the  substitution 
\begin{equation}
(p\hspace{-5pt}\slash-m_t+i\epsilon)^{-1} \to
  (p\hspace{-5pt}\slash-\mu_t+i\epsilon)^{-1}, ~~~~~~~~~~~
  \mu_t^2=m_t^2-im_t\Gamma_t.
\end{equation}
Since we are interested in NLO QCD corrections, 
gauge bosons are treated within the fixed width scheme.
 Our LO results have been generated with the
\textsc{Helac-Dipoles} \cite{Czakon:2009ss} package and cross checked
with \textsc{Helac-Phegas} \cite{Kanaki:2000ey,Cafarella:2007pc}, a
generator for  all parton level processes in the Standard Model, which
has, on its own, already been extensively used and tested in
phenomenological studies see {\it e.g.\/}
\cite{Gleisberg:2003bi,Papadopoulos:2005ky,Alwall:2007fs,
  Englert:2008tn,Actis:2010gg}.  The integration over the fractions
$x_1$ and $x_2$ of the initial partons is optimized with the help of
\textsc{Parni} \cite{vanHameren:2007pt}. The phase space integration
is executed with the help of  \textsc{Kaleu} \cite{vanHameren:2010gg}
and cross checked with \textsc{Phegas} \cite{Papadopoulos:2000tt},
both general purpose multi-channel phase space generators.  

Furthermore, results have been checked against another program  that
computes  the $t\bar{t}$ production cross section with top decays,
namely \textsc{Mcfm} \cite{mcfm}.  A perfect agreement has been found
with our results, both for the TeVatron and the LHC,  once top quarks
and $W$ gauge bosons have been put on shell in the
\textsc{Helac-Dipoles} package. We additionally reproduced results
presented  in Ref. \cite{Melnikov:2009dn} again assuming that both
tops and $W$'s are on shell. 
 
\subsection{The virtual corrections}

\begin{figure}
\begin{center}
\includegraphics[width=0.99\textwidth]{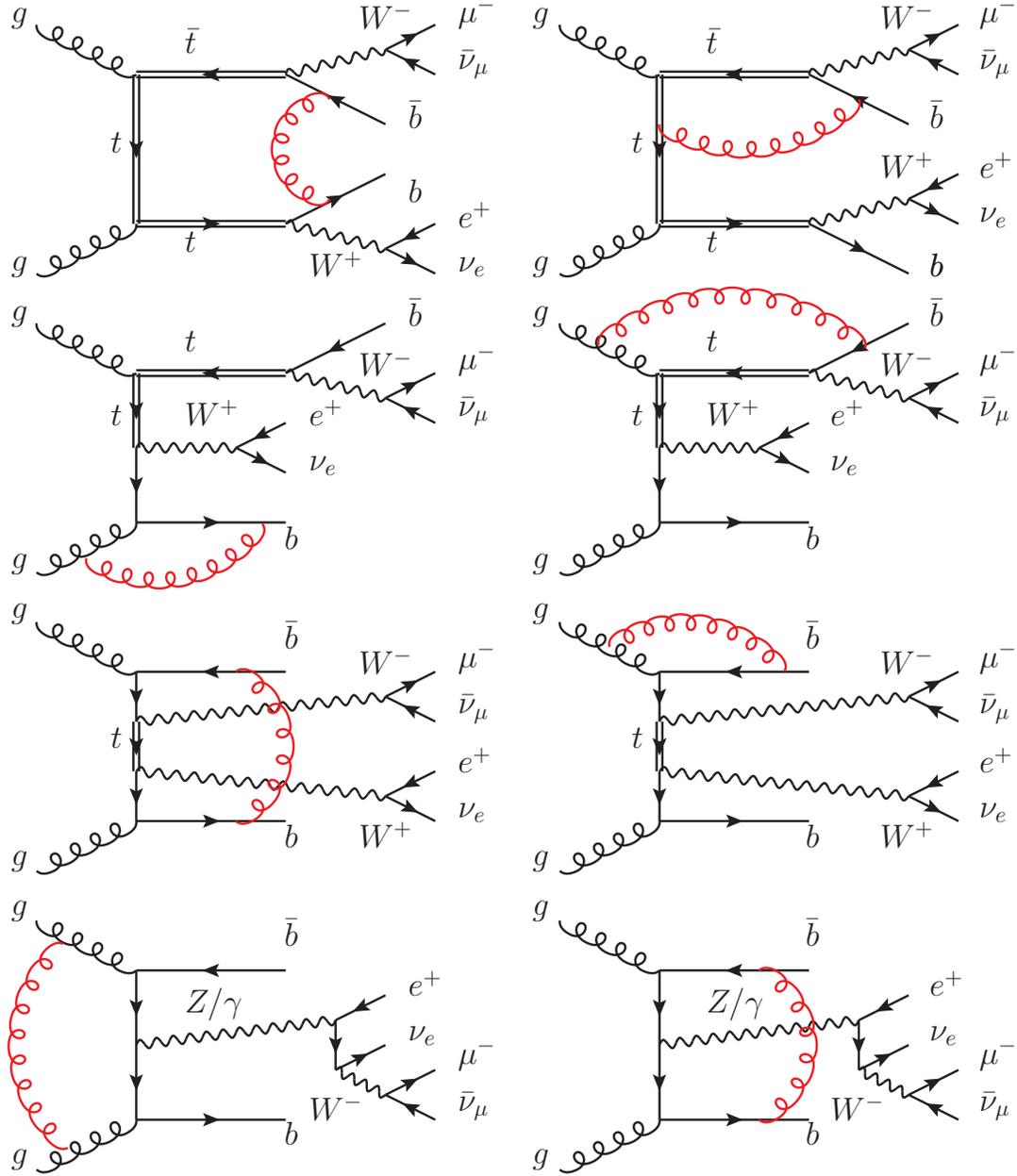}
\end{center}
\caption{\it \label{fig:diagrams-nlo}   Representative Feynman
  diagrams contributing to the  virtual corrections to the partonic
  subprocess  $gg\to  e^{+}\nu_{e}\mu^{-}\bar{\nu}_{\mu}b\bar{b}$ at
  ${\cal{O}}(\alpha_{s}^3 \alpha^4)$.}
\end{figure}

The virtual corrections consist of the 1-loop corrections to the LO
reactions.  One can classify the corrections into self-energy, vertex,
box-type,  pentagon-type and hexagon-type corrections. Typical
examples of the virtual  graphs are shown in  Figure
\ref{fig:diagrams-nlo}.  In evaluating the virtual corrections, the
\textsc{ Helac-1Loop} \cite{vanHameren:2009dr} approach is used.  It
is based on the \textsc{Helac-Phegas} program to calculate all
tree-order like ingredients and the OPP \cite{Ossola:2006us} reduction
method. The cut-constructible part of the virtual amplitudes is
computed using the \textsc{CutTools} \cite{Ossola:2007ax} code. The
rational term $R_1$ of the amplitude is computed by the
\textsc{CutTools} code as well, whereas the $R_2$ term, by the use of
extra Feynman 
rules as described in \cite{Ossola:2007ax,Draggiotis:2009yb}.
Numerical results are obtained using the same methods as described in
\cite{Bevilacqua:2009zn}.

As explained before, the process under consideration requires a
special treatment of unstable top quarks, which is achieved within the
complex-mass scheme \cite{Denner:1999gp}.  At the one-loop level the
appearance of a  non-zero top-quark width in the propagator requires
the evaluation of scalar integrals with complex masses, for which the
program \textsc{OneLOop} \cite{vanHameren:2009dr,vanHameren:2010cp} is
used.  We also need mass renormalization for the top quark, which, for
consistency, is done by using a complex mass in the well known
on-shell mass counterterm. The preservation of gauge symmetries (Ward
Identities) 
\cite{Denner:1999gp,Beenakker:1996kn,Argyres:1995ym,Dittmaier:2003bc}  by
this approach has been explicitly checked up to the one-loop level.

Although finite width effects have been studied routinely at tree
order, the same is not true for calculations at the one loop level. A
novel aspect of the introduction of a non-zero width is the effect on
the infrared structure of the scattering amplitudes. Working in
dimensional regularization, soft and collinear singularities arise.
When massive particles acquire a complex mass, the soft
$1/\epsilon$-singularities due to the exchange of gluons, are replaced
by factors proportional  to $\log(\Gamma_{t}/m_{t})$, that become
singular in the limit $\Gamma_{t}\to 0$.  We have explicitly checked
that including all contributions, factorizable and non-factorizable,
the usual cancellation of infrared $1/\epsilon^2$ and $1/\epsilon$
poles between virtual and real corrections, the latter represented by
the ${\cal{I}}(\epsilon)$-operator, takes place. This means that a
partial cancellation of $\log(\Gamma_{t}/m_{t})$ terms happens within
the virtual corrections alone.  Nevertheless logarithmic enhancements
remain in the finite part of the virtual corrections and have to be
cancelled by corresponding terms from the real corrections, since they
represent the same soft singularities, dimensionally regularized in
the case of on-shell particles.

\subsection{The real emission}

The generic processes for the real corrections are given by 
\[
gg\to e^{+}\nu_{e}\mu^{-}\bar{\nu}_{\mu}b\bar{b}g
\]
\[
qg\to e^{+}\nu_{e}\mu^{-}\bar{\nu}_{\mu}b\bar{b} q
\]
\begin{equation}
gq\to e^{+}\nu_{e}\mu^{-}\bar{\nu}_{\mu}b\bar{b}q
\end{equation}
\[
q\bar{q}\to e^{+}\nu_{e}\mu^{-}\bar{\nu}_{\mu}b\bar{b} g 
\]
(where again $q$ stands for up- or down-type quarks) and include all
possible contributions of the order of ${\cal{O}}(\alpha_{s}^3
\alpha^4)$. The complex mass   scheme for unstable top quarks has been
implemented in complete analogy to the  LO case.

We employ the dipole subtraction formalism \cite{Catani:1996vz} to
extract the soft and collinear infrared singularities and to combine
them with the virtual corrections.  Specifically, the formulation
\cite{Catani:2002hc} for massive quarks has been used with the
extension to arbitrary helicity   eigenstates of the external partons
\cite{Czakon:2009ss},  as implemented in \textsc{Helac-Dipoles}. In
the case at hand, the number of dipoles is as  follows: $27$ for the
process $gg\to e^{+}\nu_{e}\mu^{-}\bar{\nu}_{\mu}b\bar{b}g$  and  $15$
for processes $q\bar{q}\to e^{+}\nu_{e}\mu^{-}\bar{\nu}_{\mu}b\bar{b}
g$,  $qg\to e^{+}\nu_{e}\mu^{-}\bar{\nu}_{\mu}b\bar{b}q$ and  $gq\to
e^{+}\nu_{e}\mu^{-}\bar{\nu}_{\mu}b\bar{b}q$.  Let us stress at this
point, that, similarly to most authors, we do not use finite dipoles
regularizing the quasi-collinear divergence induced by both top quarks
moving in the same direction, even though they are implemented in the
software. Due to the large top quark mass, they do not improve
numerical stability.

Besides the cancellation of divergences, which we have mentioned in
the previous section, we have also explored the independence of  the
results on the unphysical cutoff in the dipole subtraction phase space
(see \cite{Czakon:2009ss} and references therein for details) to
further check our calculation.
 
\subsection{Phase space generation}

In LO calculations, the jet definition consists of a set of phase
space cuts not allowing any parton to become arbitrarily soft, and no
pair of partons to become arbitrarily collinear.  This changes for the
real-radiation contribution in NLO calculations, for which single
partons are allowed to become arbitrarily soft and single pairs of
partons are allowed to become arbitrarily collinear.  This means that
phase space generators like \textsc{Phegas} \cite{Papadopoulos:2000tt}
and \textsc{Kaleu} \cite{vanHameren:2010gg}, which construct momentum
configurations from kinematical variables generated following {\em a
  priori\/} defined probability densities, cannot be directly applied
in their LO set-up, since these densities anticipate the singular
behavior of the squared amplitudes, and are typically not defined in
the soft and collinear limits.  Furthermore, the subtraction terms in
the dipole-subtraction scheme, used to eliminate the singularities in
the real-radiation phase space integral, do not exactly follow the
same peak structure as the tree-level $\nparton+1$-particle matrix
element squared, whereas \textsc{Phegas} and \textsc{Kaleu}  are
designed only  to efficiently deal with the latter.  We chose to deal
with this situation via a multi-channel approach \cite{Kleiss:1994qy},
in which a separate channel is associated with each term in the
real-subtracted integral, {\it i.e.}, with the tree-level
$\nparton+1$-particle matrix element squared as well as each dipole
term.

The channel for the $\nparton+1$-particle matrix element squared
generates momenta using an instance of \textsc{Kaleu}  anticipating
the peak structure of this integrand.  The phase space defined by
promoting the LO cuts to $\nparton+1$ partons is filled in the usual
LO approach.  The soft and collinear regions ``below the cuts'' are
filled by replacing the densities for the invariants by densities that
are integrable in these regions.

All dipole channels also carry their own instances of \textsc{Kaleu},
but each of these generates $\nparton$-momentum configurations
anticipating the peak structure of the $\nparton$-particle matrix
element squared of the underlying process of the dipole term.  Such a
$\nparton$-momentum configuration is then turned into an
$\nparton+1$-momentum configuration by essentially applying the
inverse of the phase space mapping performed in the calculation of the
dipole contribution itself.  This generation of an extra momentum
follows exactly the formulas for the parton showers based on the
dipole formalism presented in \cite{Dinsdale:2007mf} and
\cite{Schumann:2007mg}.   The azimuthal angle needed for the
construction of the extra momentum is generated with a flat
distribution, and the other two variables, traditionally denoted
$(y_{ij,k},z_i)$ for final-final, $(x_{ij,a},z_i)$ for final-initial,
$(x_{ij,a},u_i)$ for initial-final, and $(x_{i,ab},v_i)$ for
initial-initial dipoles, are generated following  self-adaptive
densities.  This happens ``on the fly'' during the Monte Carlo
integration, following the approach presented in
\cite{vanHameren:2007pt}.  Finally, each instance of \textsc{Kaleu}
carries a multi-channel weight in the ``highest level'' multi-channel
density which is optimized during the Monte Carlo integration, and
each instance performs its own internal multi-channel optimization, as
described in~\cite{vanHameren:2010gg}.

We have performed a few tests to check  the performance of this new
approach in case of the $q\bar{q}\to
e^{+}\nu_{e}\mu^{-}\bar{\nu}_{\mu}b\bar{b} g$ subprocess. More
precisely, we have made a comparison between three options, namely
\textsc{Kaleu} with dipole channels, \textsc{Kaleu} without dipole
channels and \textsc{Phegas}, which does not have dipole
channels. Since the computational cost comes mainly from the accepted
events, comparisons are made at equal numbers of accepted events.  Our
findings can be summarized as follows.    For the dipole phase space
cut-off parameter $\alpha_{max} = 1$, when all dipoles are calculated
for each phase space point, \textsc{Phegas} and \textsc{Kaleu} without
dipoles channels are comparable in terms of errors. \textsc{Kaleu}
with dipole channels, however, gives an error which is 5 times
smaller. Realize that this implies a reduction in the number of events
by a factor of 25 to reach the same error.  For $\alpha_{max} = 0.01$,
when much less dipole subtraction terms are  needed per  event, the
improvement is not so dramatic, and the introduction of the dipole
channels reduces the error by a factor 3 compared to \textsc{Kaleu}
without dipole channels, and a factor 2 compared to \textsc{Phegas},
implying a reduction in necessary events by a factor of 9 and 4
respectively.

We conclude that the dipole channels structurally improve the
convergence of the phase space integrals for the real-subtracted
contribution. It is, however, difficult to express the improvement
quantitatively because it depends on the process and the value of
parameters like $\alpha_{max}$.
  
\section{Numerical Results}

\label{sec:results}

\subsection{Setup}
\label{sec:setup}

We consider the process $pp(p\bar{p}) \rightarrow  t\bar{t} + X
\rightarrow  W^+W^-b\bar{b} + X \rightarrow e^+ \nu_e \mu^- \nu_{\mu}
b\bar{b} +X$  both at the TeVatron run II and the LHC {\it i.e.}  at a
center-of-mass energy of $\sqrt{s} = 1.96$ TeV and $\sqrt{s} = 7$ TeV
correspondingly.  For the LHC case we additionally calculate the
integrated cross  section at a center-of-mass energy  $\sqrt{s} = 10$
TeV.  We only simulate decays of the weak  bosons to different lepton
generations to avoid  virtual photon singularities stemming from
quasi-collinear $\gamma^{*}\rightarrow \ell^+\ell^-$ decays.  These
interference effects are at the per-mille level for inclusive cuts, as
checked by an explicit leading order calculation. The complete
$\ell_1^\pm\ell_2^\mp$ cross section (with  $\ell_{1,2}=e,\mu$)  can
be obtained by multiplying the result with a lepton-flavor factor of
4.  We keep the Cabibbo-Kobayashi-Maskawa mixing matrix diagonal.  The
unstable (anti)top quark is treated within the (gauge invariant)
complex-mass scheme, as explained in the previous  section.  The
Standard Model parameters are given the following values within the
$G_\mu$ scheme\cite{Amsler:2008zzb}: 
\[
m_W = 80.398 ~\textnormal{GeV}, ~~~~ \Gamma_{W}=2.141 ~\textnormal{GeV},
\]
\[
m_Z=91.1876  ~\textnormal{GeV}, ~~~~ \Gamma_Z=2.4952 ~\textnormal{GeV},
\]
\[
G_\mu = 1.16639 \times 10^{-5} ~\textnormal{GeV}^{-2},
\]
\begin{equation}
 \sin^2\theta_W	= 1 - m^2_W /m^2_Z.
\end{equation}
The electromagnetic coupling is derived from the Fermi constant $G_\mu$ 
according to
\begin{equation}
\alpha = \frac{\sqrt{2}G_{\mu}m_W^2 \sin^2\theta_W}{\pi} ~.
\end{equation}
For the top quark mass we take   $m_t = 172.6 $ GeV and  all other QCD
partons including $b$ quarks as well as leptons  are treated as
massless. The contribution from the Higgs boson can be  neglected
since for inclusive cuts  it is below $1 \%$. In our case, however,
the $b$-quarks are massless and   the Higgs contribution simply
vanishes.  The top quark width calculated from
\cite{Jezabek:1988iv,Chetyrkin:1999ju}  is $\Gamma_{t}^{LO}=1.48
~\mbox{GeV}$ at LO and $\Gamma_{t}^{NLO}=1.35 ~\mbox{GeV}$ at NLO
where $\alpha_s = \alpha_s(m_{t})= 0.107639510785815$.  Mass
renormalization is performed in the on-shell scheme.  All final-state
$b$ quarks and gluons with pseudorapidity $|\eta| < 5$
are  recombined into jets with
separation $\sqrt{\Delta\phi^2 +\Delta y^2} > D  = 0.4$ in the
rapidity-azimuthal angle plane via the following IR-safe algorithmes: the $k_T$
algorithm \cite{Catani:1992zp,Catani:1993hr,Ellis:1993tq},  the {\it
  anti-}$k_T$ algorithm \cite{Cacciari:2008gp} and the inclusive
Cambridge/Aachen algorithm (C/A) \cite{Dokshitzer:1997in}.
The distance measure $d_{ij}$ for these algorithms is defined as
\[
d_{ij}= \min \left(p_{T}^{2p}(i),p_{T}^{2p}(j) \right)\frac{\Delta R^2_{ij}}{D^2}
\]
\[
d_{iB}= p_{T}^{2p}(i) ~,
\]
where $\Delta R_{ij}= \sqrt{\Delta \phi_{ij}^2 + \Delta y_{ij}^2}$ and
the parameter $p$ is equal to $1$ for the $k_T$ algorithm,  $0$ for
$C/A$ and $-1$ for {\it anti}-$k_T$ algorithm. Moreover, we impose
the  following additional cuts on the transverse momenta and the
rapidity of two  recombined $b$-jets: $p_{T_{b}} > 20$ GeV, 
$|y_b| < 4.5$ where 
\[
p_T= \sqrt{p_x^2+p_y^2}, 
~~~~~y = \frac{1}{2}\ln\left(\frac{E+p_{z}}{E-p_{z}}\right).
\]
Basic selection  is applied to (anti)top decay products to ensure that
the leptons are observed inside the detector and are well separated
from each other: $p_{T_{\ell}}> 20 ~\textnormal{GeV}$, $|\eta_\ell| <
2.5$, $\Delta R_{j\ell} > 0.4$, where $j=b,\bar{b}$, and  $p_{T_{miss}} > 30
~\textnormal{GeV}$.  In the following we consistently use the CTEQ6
set of parton  distribution functions (PDFs)
\cite{Pumplin:2002vw,Stump:2003yu}. More precisely,  we take CTEQ6L1
PDFs with a 1-loop running $\alpha_s$ in LO and CTEQ6M PDFs with a
2-loop running $\alpha_s$ in NLO. The contribution from $b$ quarks in
the initial state is neglected, since at LO for inclusive cuts this
contribution is suppressed to the per-mille level. The number of
active flavors is $N_F = 5$, and the respective QCD parameters are
$\Lambda^{LO}_5 = 165$ MeV and  $\Lambda^{MS}_5 = 226$ MeV. In the
renormalization of the strong coupling constant, the top-quark loop in
the gluon self-energy is subtracted at zero momentum. In this scheme
the running of $\alpha_s$ is generated solely by the contributions of
the light-quark and gluon loops. By default, we set the
renormalization and factorization scales, $\mu_{R}$  and $\mu_F$, to
the  common value $\mu =m_t$.  For inclusive cuts, where the
contribution from the double resonance Feynman diagrams dominates, the
top mass is a  valid scale.

\subsection{Results for the TeVatron run II}
\label{sec:tev}

We begin our presentation of the final results of our analysis with a
discussion of the total cross section at the central value of the
scale, $\mu_R=\mu_F=m_t$ at the TeVatron run II.   The respective
numbers are presented in Table \ref{tab:tev} for the two choices of
the dipole phase space cutoff    parameter $\alpha_{max}$ (see {\it
  e.g.} \cite{Czakon:2009ss} for more details) and for three different
jet algorithms.  At the central scale value, the full cross section
receives small NLO correction of the order of $2.3\%$.

\begin{figure}[th]
\begin{center}
\includegraphics[width=0.49\textwidth]{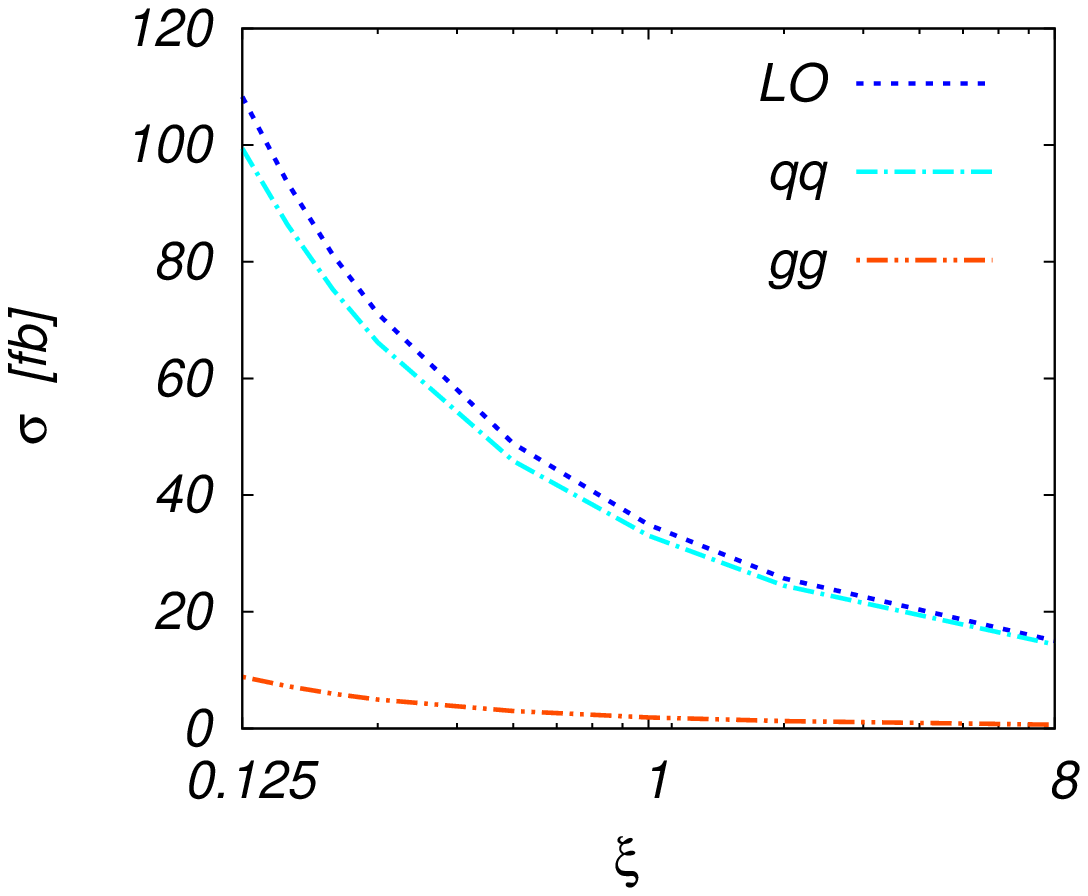}
\includegraphics[width=0.49\textwidth]{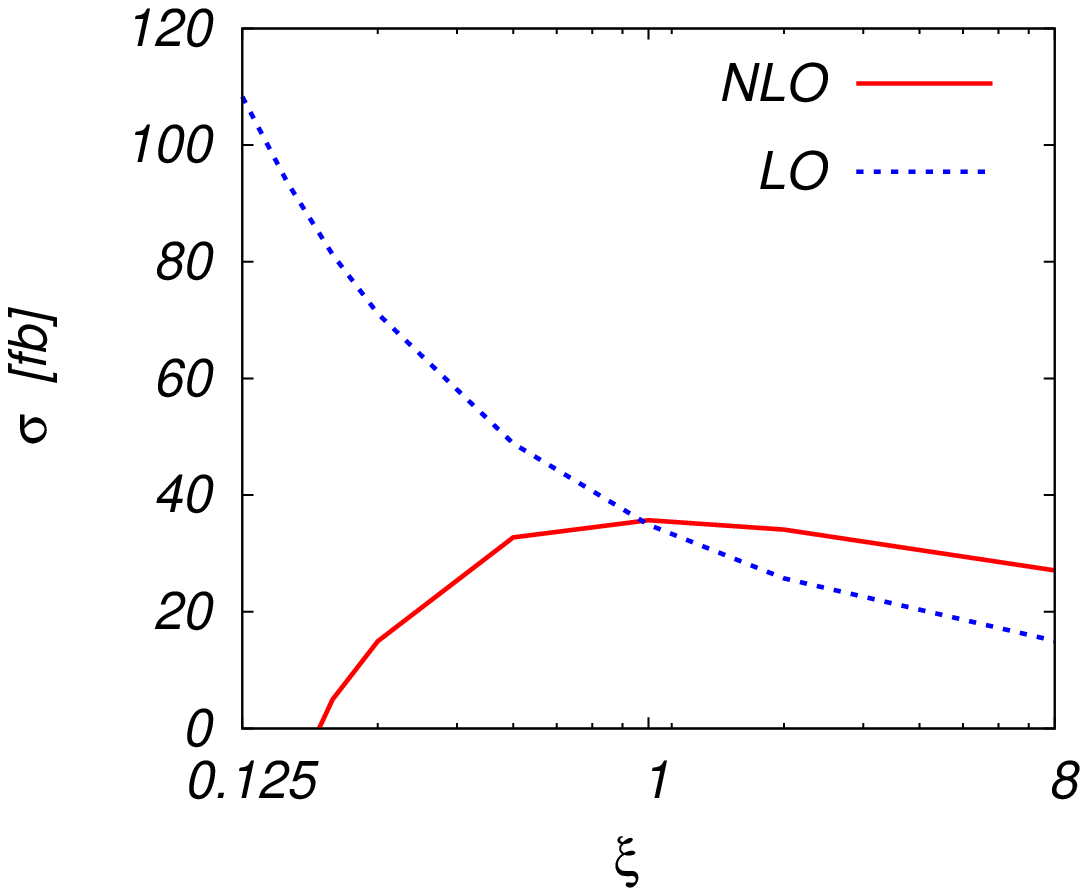}
\end{center}
\caption{\it \label{fig:scales-tev} Scale dependence of the LO cross
  section with the individual contributions of the partonic channels
  (left panel) and  scale dependence of the LO and NLO cross sections
  (right panel)  for the  $p\bar{p}\rightarrow
  e^{+}\nu_{e}\mu^{-}\bar{\nu}_{\mu}b\bar{b} ~+ X$ process at the
  TeVatron run II with  $\sqrt{s}=1.96$ TeV, where  renormalization
  and factorization  scales are set to the common value
  $\mu=\mu_R=\mu_F=\xi m_t$.}
\end{figure}

Subsequently, we turn our attention to the scale dependence for the
total cross section at LO and NLO. The left panel of Figure
\ref{fig:scales-tev} shows the dependence of  the integrated LO cross
section on the renormalization and factorization  scales where
$\mu=\mu_R=\mu_F=\xi m_t$. The variation range is taken from  $\mu=m_t/8$ to
$\mu=8 m_t$.  The dependence is large, illustrating the well known
fact  that the LO prediction can only provide  a rough estimate. At
the TeVatron  with our cut selection the $q\bar{q}$ channel dominates
the total $p\bar{p}$  cross section by about $95 \%$ followed by the
$gg$ channel with about $5\%$.  In the right panel the scale
dependence of the NLO cross section is shown  together with the LO
one.  As expected, we observe a reduction of the scale uncertainty
while going from LO to NLO. Varying the scale down and up by a factor
2 changes the cross section by $+40\%$ and $-26\%$ in the LO case,
while  in the NLO case we have obtained a variation of the order
$-8\%$ and $-4\%$.   Let us mention here that while calculating the
scale dependence for the NLO  cross section we kept $\Gamma^{NLO}_{t}$
fixed independently  of the scale  choice. The error introduced by
this treatment is however of higher order, and particularly for two
scales $\mu=m_t/2$ and $\mu=2 m_t$   amounts to $\pm 1.5\%$
respectively.

\begin{table}[th]
\begin{center}
  \begin{tabular}{|c|c|c|c|}
    \hline
      && & \\
      Algorithm & $\sigma_{\rm LO}$ [fb]      & 
     $\sigma_{\rm NLO}^{\rm \alpha_{max}=1}$  [fb] &   
      $\sigma_{\rm NLO}^{\rm \alpha_{max}=0.01}$   [fb] \\
      && & \\
    \hline
{\it anti}-$k_T$ 
& 34.922 $\pm$ 0.014 & 35.705 $\pm$ 0.047 & 35.697 $\pm$ 0.049 \\
    \hline 
$k_T$ &  34.922 $\pm$ 0.014 & 35.727 $\pm$ 0.047 & 35.723 $\pm$  0.049   \\
 \hline
C/A &  34.922 $\pm$ 0.014 & 35.724 $\pm$ 0.047 & 35.746  $\pm$ 0.050    \\
\hline
  \end{tabular}
\end{center}
  \caption{\it \label{tab:tev} Integrated cross section at LO and NLO
    for  $p\bar{p}\rightarrow
    e^{+}\nu_{e}\mu^{-}\bar{\nu}_{\mu}b\bar{b} ~+ X$  production at
    the TeVatron run II with $\sqrt{s}= 1.96  ~\textnormal{TeV}$,  for
    three different jet algorithms,  the anti-$k_T$, $k_T$ and the
    Cambridge/Aachen jet algorithm.  The two NLO results refer to
    different values of  the dipole phase space cutoff
    $\alpha_{max}$. The scale choice is  $\mu_R=\mu_F=m_{t}$.}
\end{table}
\begin{figure}[th]
\begin{center}
\includegraphics[width=0.49\textwidth]{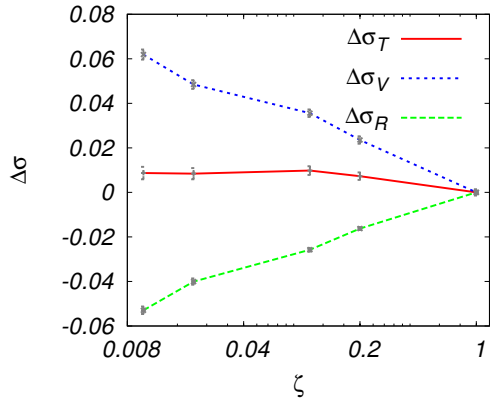}
\end{center}
\caption{\it \label{fig:rescaling-tev} Dependence of the NLO   cross
  section, $\sigma_{\rm T}$, (red solid line) and the individual
  contributions, the real emission part, $\sigma_{\rm R}$,  (green
  dashed line)  and the LO plus virtual part, $\sigma_{\rm V}$, (blue
  dotted line),  on the rescaling parameter $\zeta$ defined as
  $\Gamma_{rescaled}=  \zeta \Gamma_t$ for the
  $p\bar{p}\rightarrow e^{+}\nu_{e}\mu^{-}\bar{\nu}_{\mu}b\bar{b} ~+
  X$ process at the TeVatron run II with  $\sqrt{s}=1.96$ TeV.
  $\Delta\sigma$ is defined as follows:
  $\Delta\sigma_i(\zeta)=(\sigma_{i}(\zeta)-\sigma_{i}
  (\zeta=1))/\sigma_{\rm T}(\zeta=1)$ with $i=V,R,T$.} 
\end{figure}
\begin{figure}[th]
\begin{center}
\includegraphics[width=0.49\textwidth]{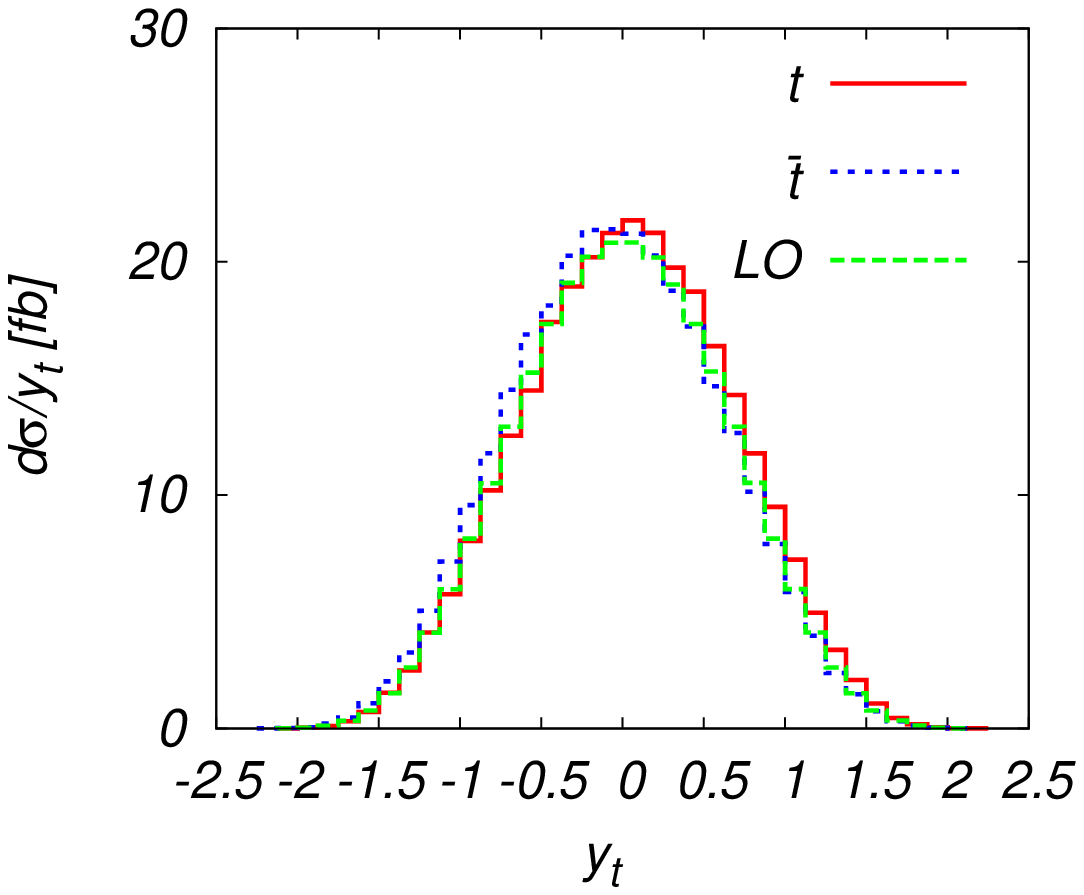}
\includegraphics[width=0.49\textwidth]{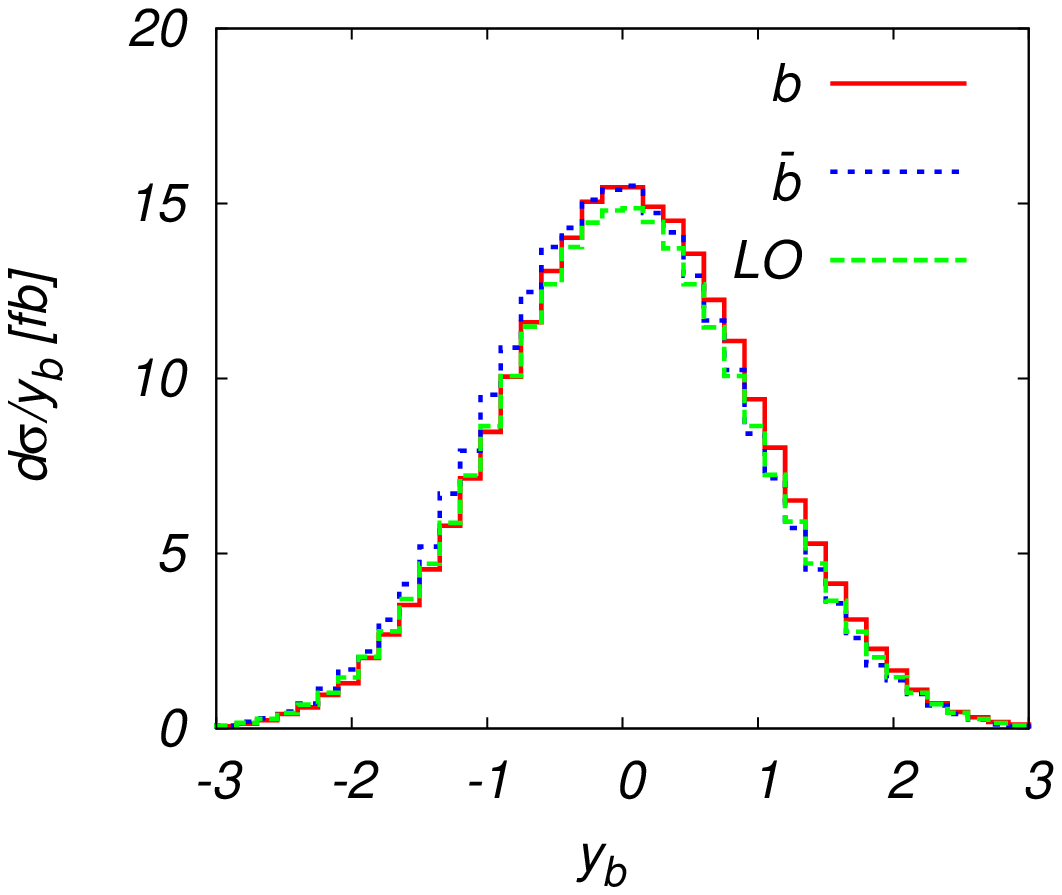}
\includegraphics[width=0.49\textwidth]{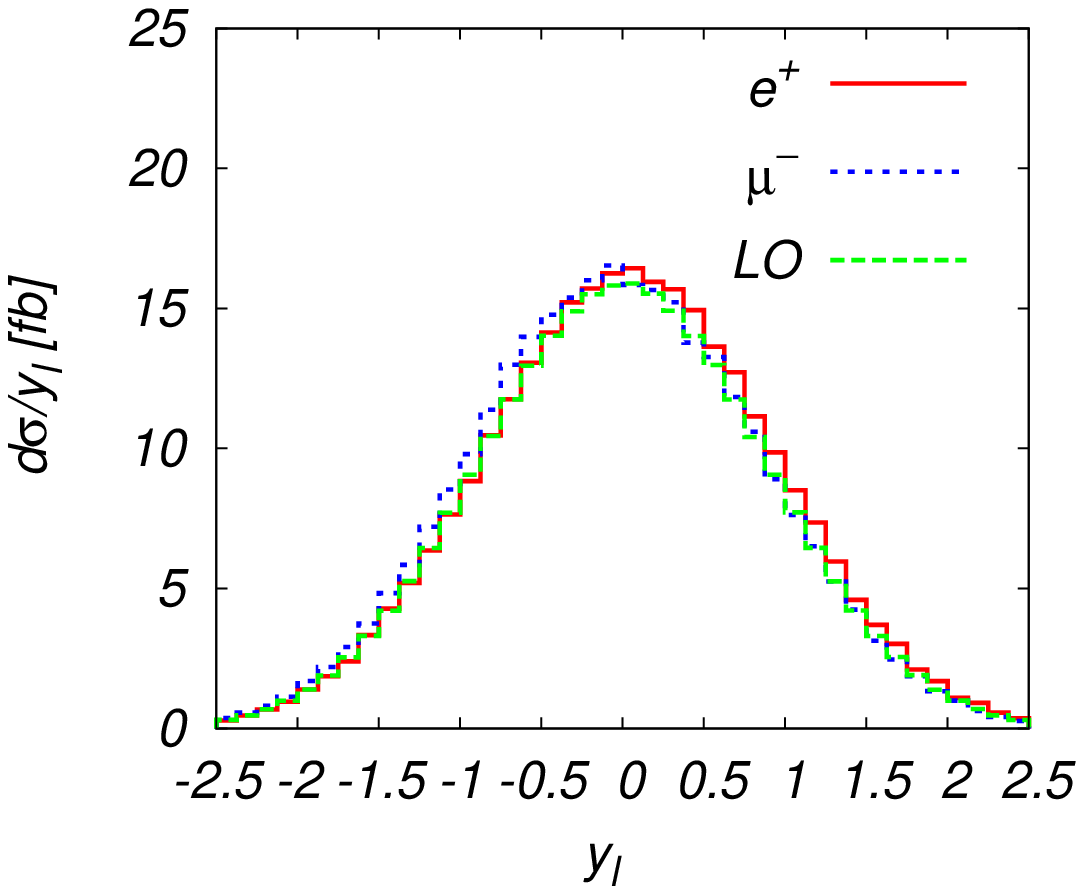}
\end{center}
\caption{\it \label{fig:asymmetry} Differential cross section
  distributions as a function of rapidity $y_{t}$  of the top  (red
  solid curve) and anti-top quarks (blue dotted curve),   rapidity
  $y_{b}$ of the b-jet (red solid curve) and anti-b-jet   (blue dotted
  curve) and rapidity $y_{l}$ of the positron  (red solid curve) and
  muon (blue dotted curve) at next-to-leading order  for the
  $pp\rightarrow  e^{+}\nu_{e}\mu^{-}\bar{\nu}_{\mu}b\bar{b} ~ + X$
  process at the TeVatron run II.  The green dashed curves correspond
  to the leading order results.}
\end{figure}
\begin{figure}[th]
\begin{center}
\includegraphics[width=0.49\textwidth]{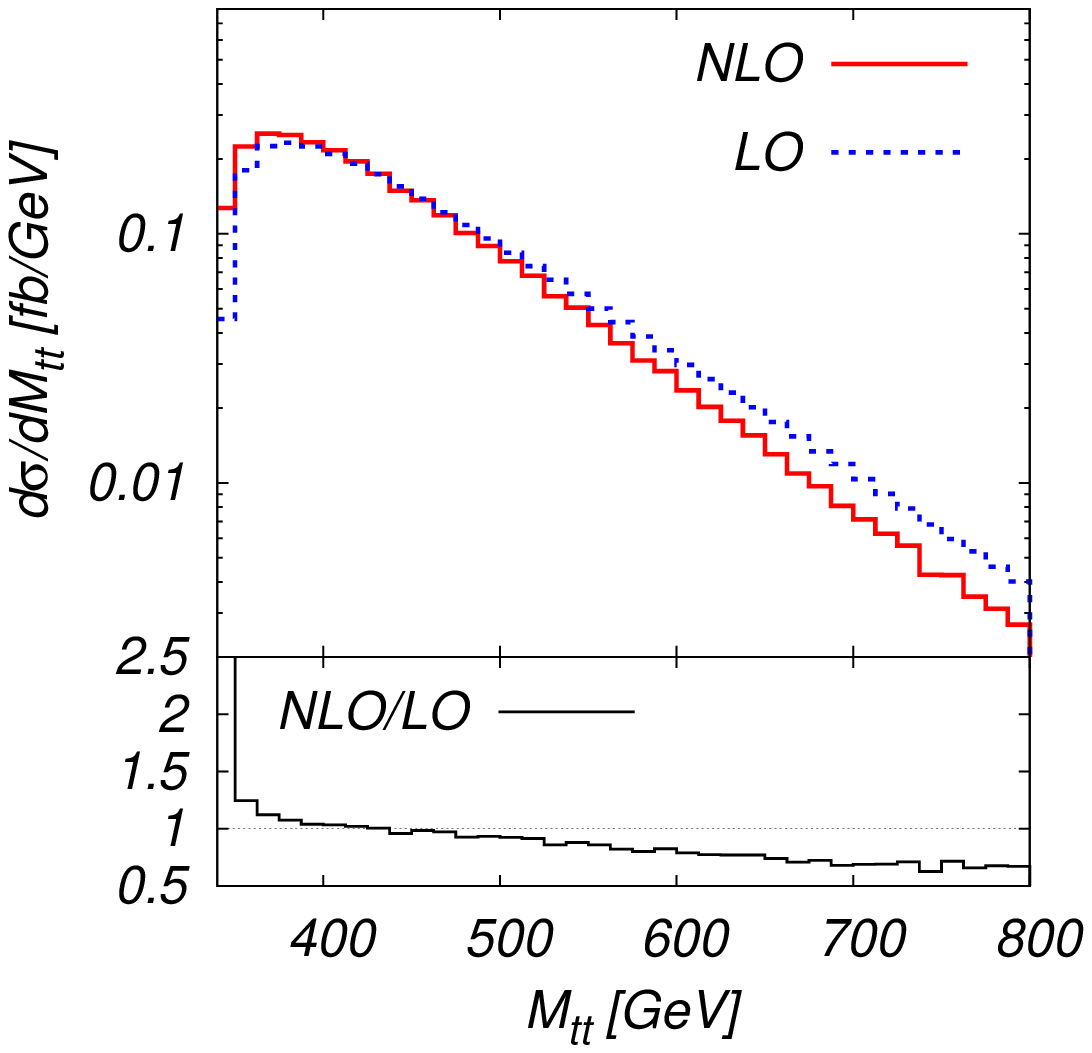}
\includegraphics[width=0.49\textwidth]{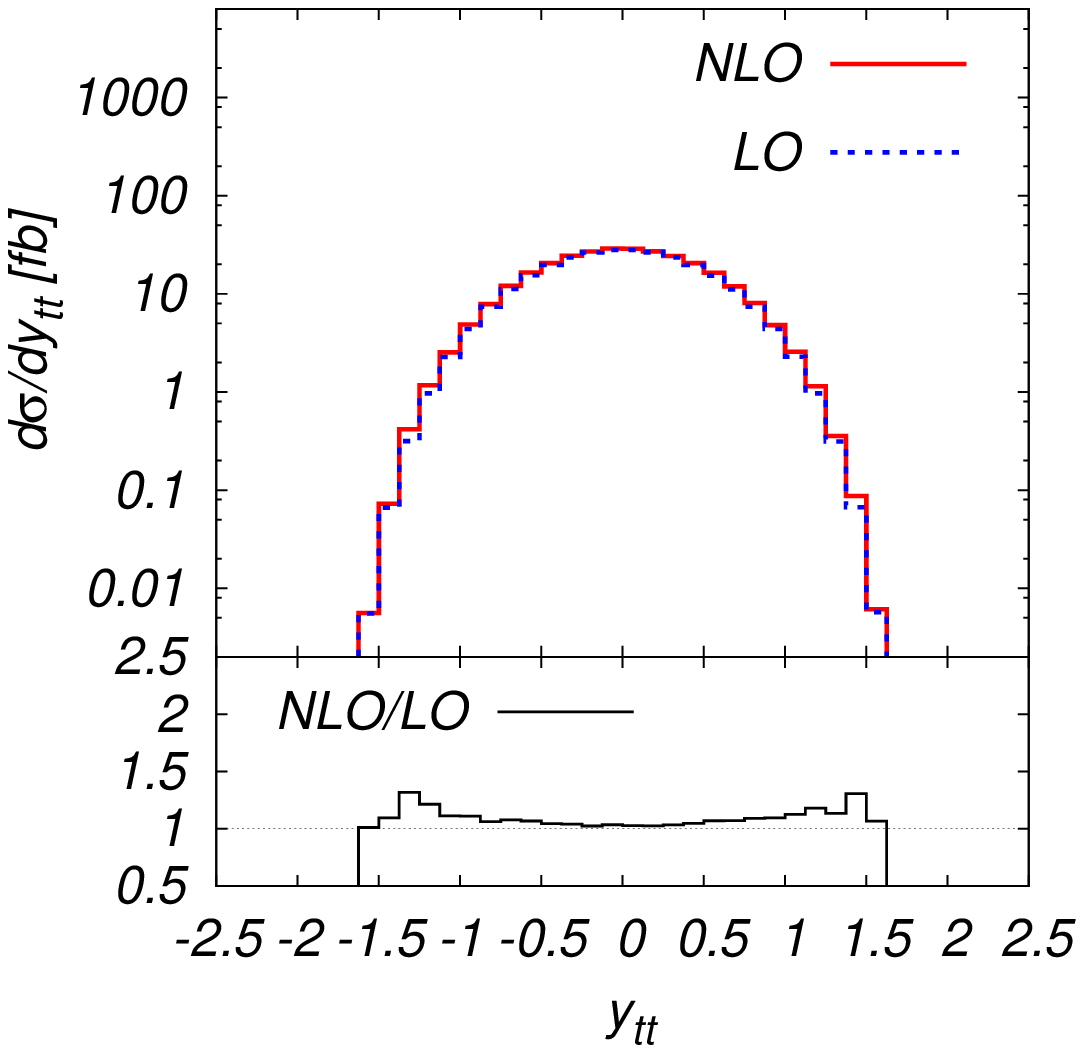}
\includegraphics[width=0.49\textwidth]{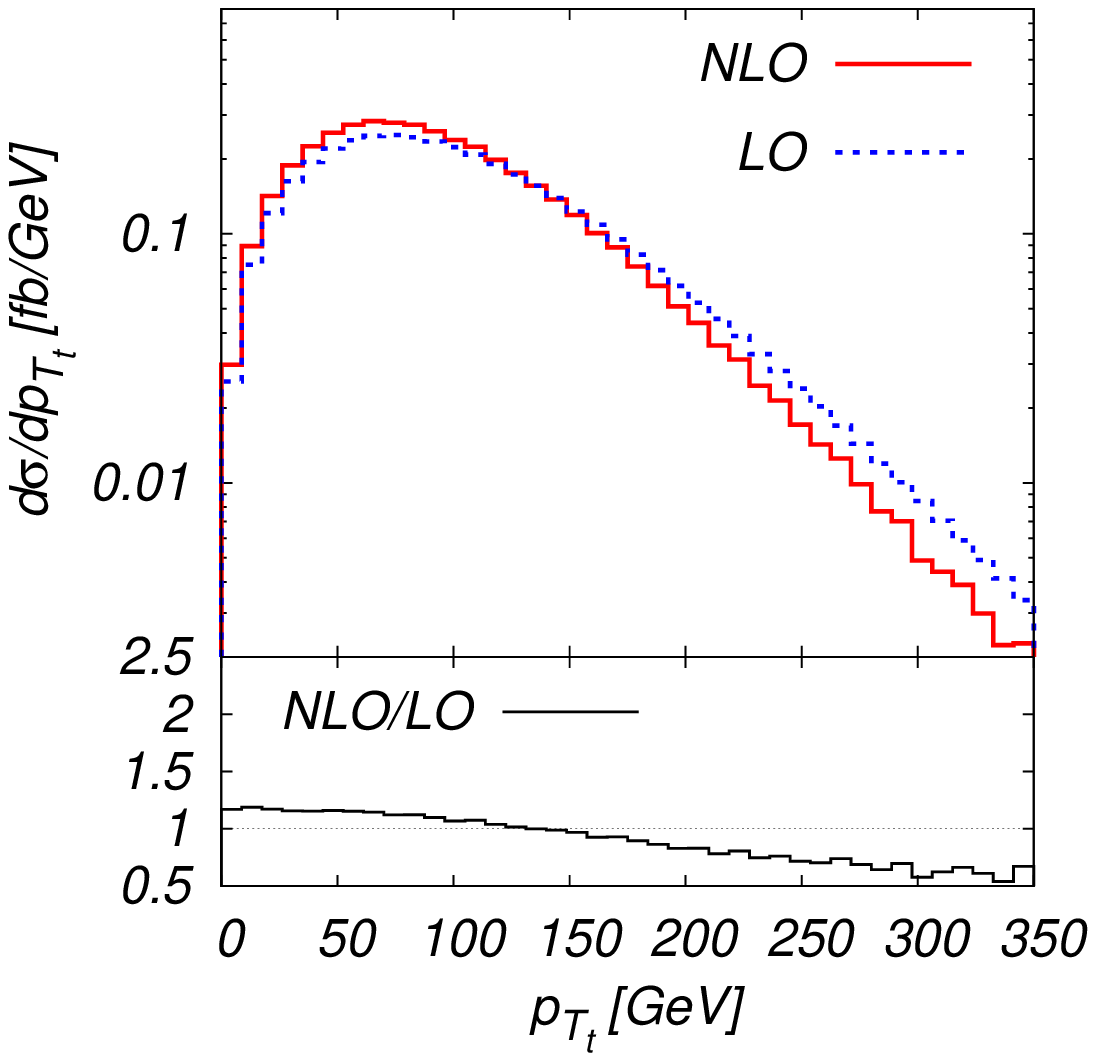}
\includegraphics[width=0.49\textwidth]{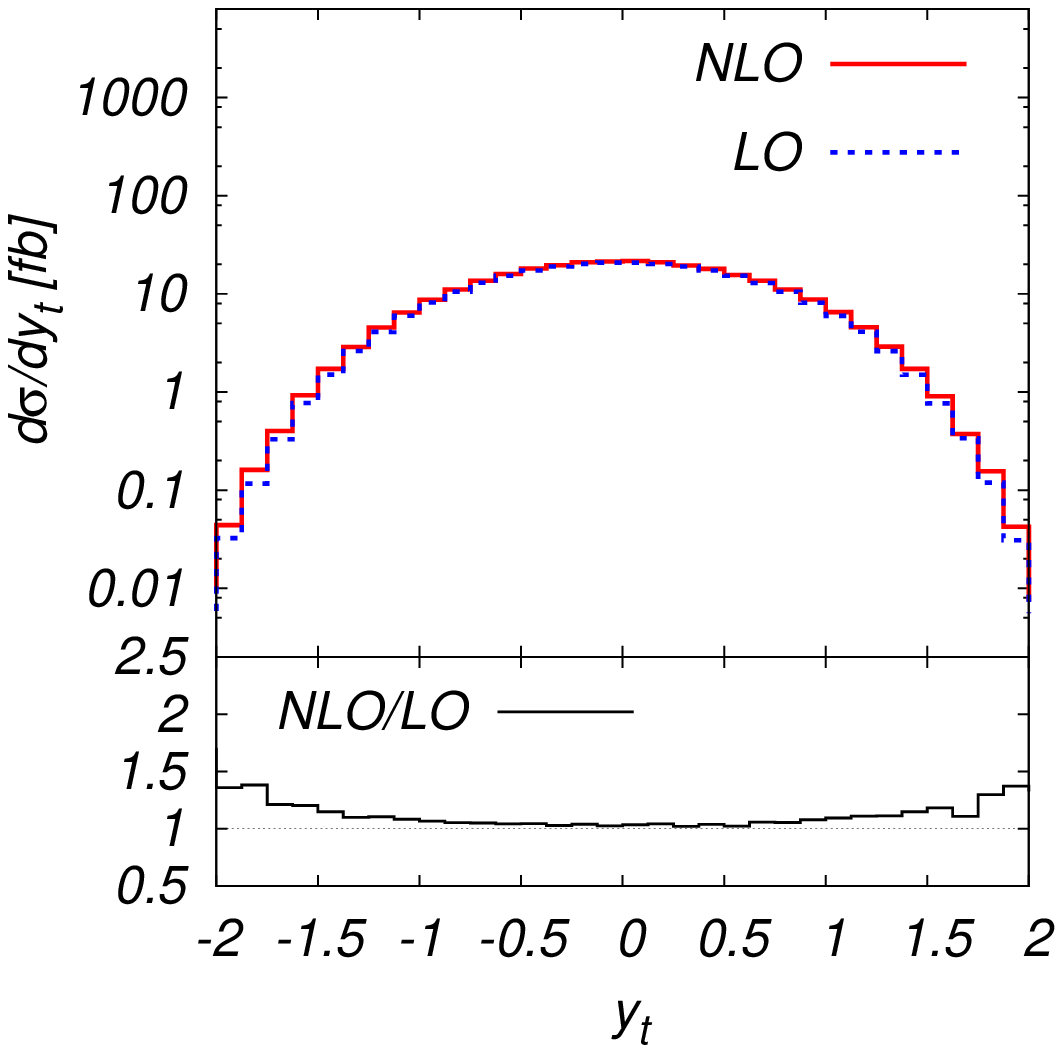}
\end{center}
\caption{\it \label{fig:top-tev} Differential  cross section
  distributions as a function  of the invariant mass $m_{t\bar{t}}$ of
  the top-anti-top pair, rapidity $y_{t\bar{t}}$ of the top-anti-top
  pair, averaged transverse momentum $p_{T_{t}}$  of the top and
  anti-top and  averaged  rapidity $y_{t}$ of  the top and anti-top
  for the $p\bar{p}\rightarrow
  e^{+}\nu_{e}\mu^{-}\bar{\nu}_{\mu}b\bar{b} ~ + X$ process at the
  TeVatron  run II.  The blue dashed curve corresponds to the leading
  order, whereas the red solid one to the next-to-leading order
  result. The lower panels display  the differential K factor.}
\end{figure}
\begin{figure}[th]
\begin{center}
\includegraphics[width=0.49\textwidth]{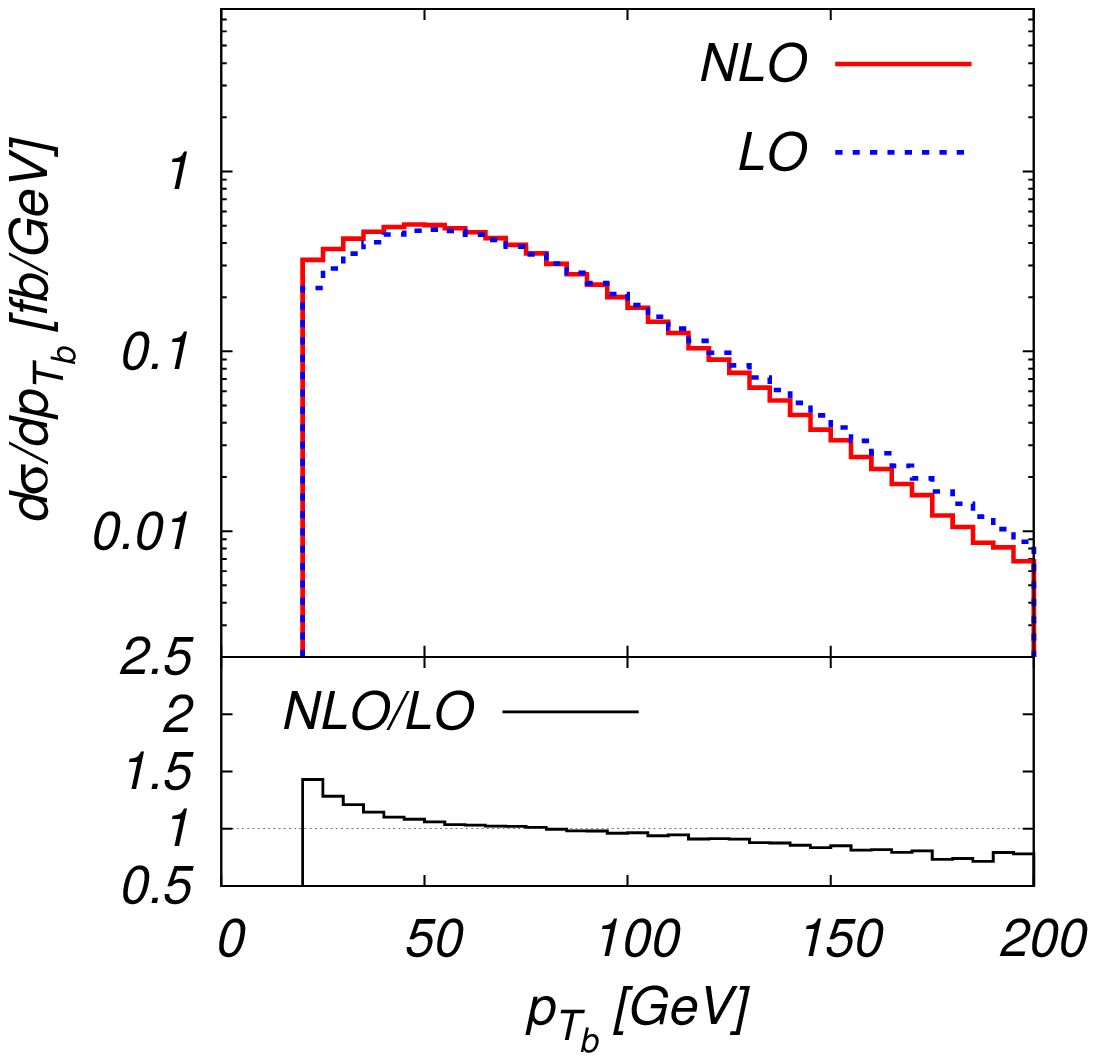}
\includegraphics[width=0.49\textwidth]{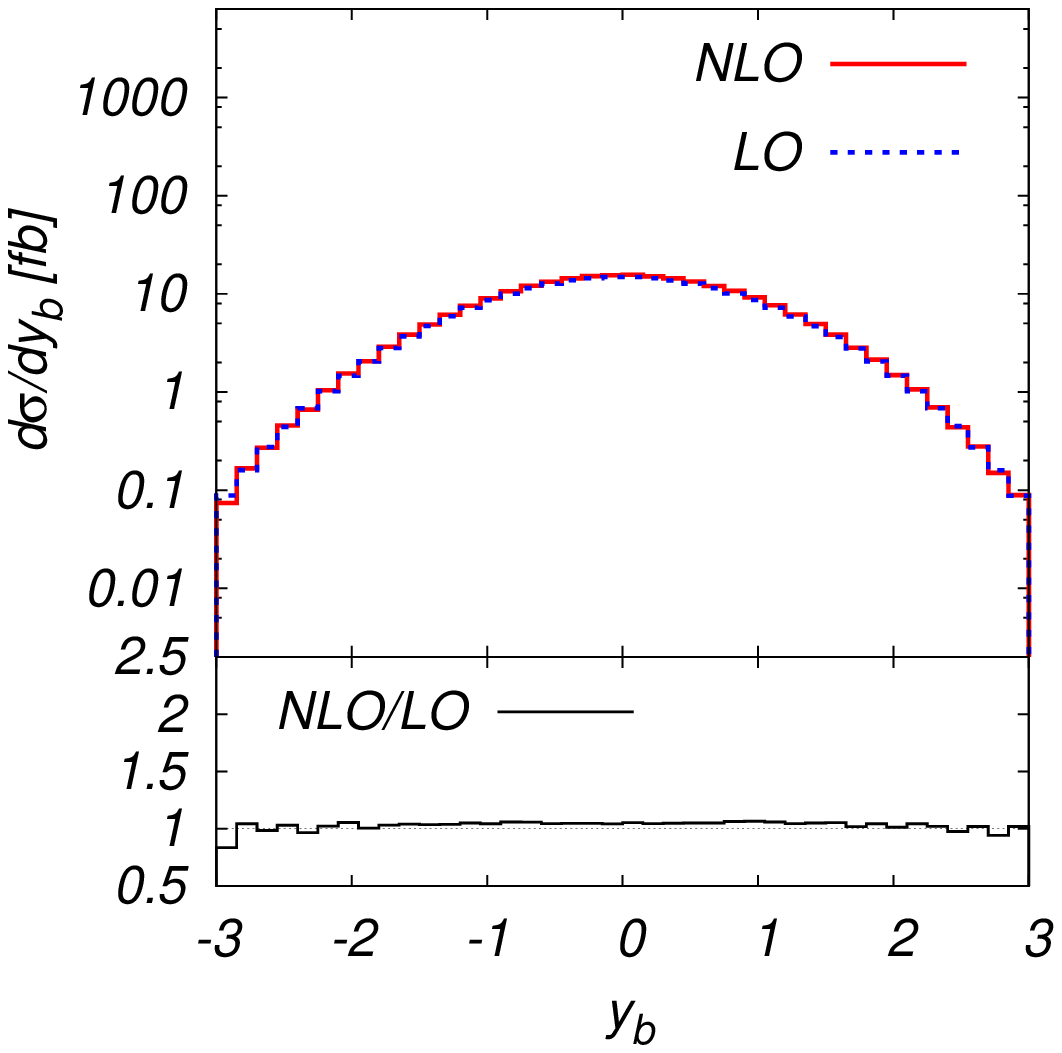}
\includegraphics[width=0.49\textwidth]{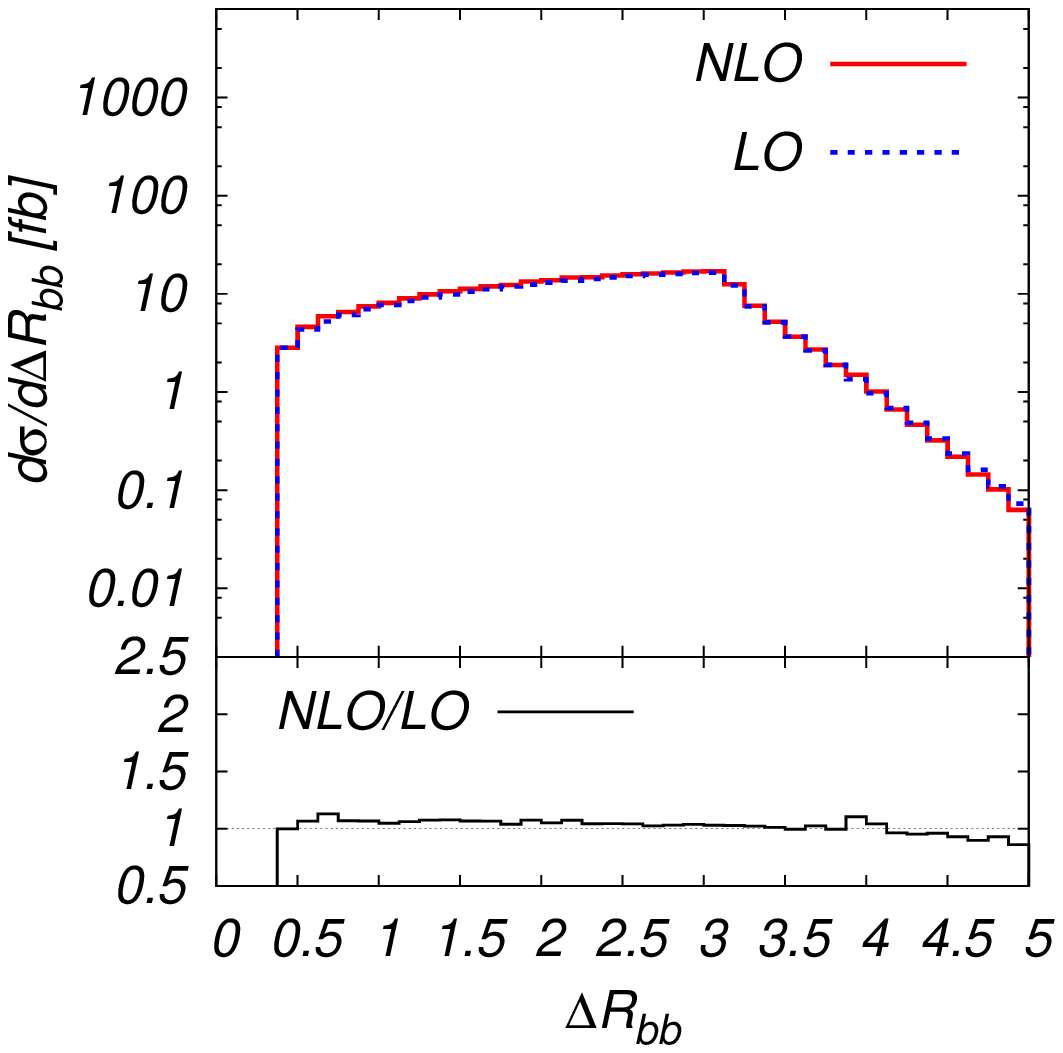}
\end{center}
\caption{\it \label{fig:bottom-tev}  Differential  cross section
  distributions as a function of the averaged  transverse momentum
  $p_{T_{b}}$  of the  b-jet and anti-b-jet,  averaged rapidity
  $y_{b}$  of the  b-jet and anti-b-jet and $\Delta R_{b\bar{b}}$
  separation for the $p\bar{p}\rightarrow
  e^{+}\nu_{e}\mu^{-}\bar{\nu}_{\mu}b\bar{b} ~ + X$ process at the
  TeVatron run II.   The blue dashed curve corresponds to the leading
  order, whereas the red solid one to the next-to-leading order
  result.  The lower panels display  the differential K factor.}
\end{figure}
\begin{figure}[th]
\begin{center}
\includegraphics[width=0.49\textwidth]{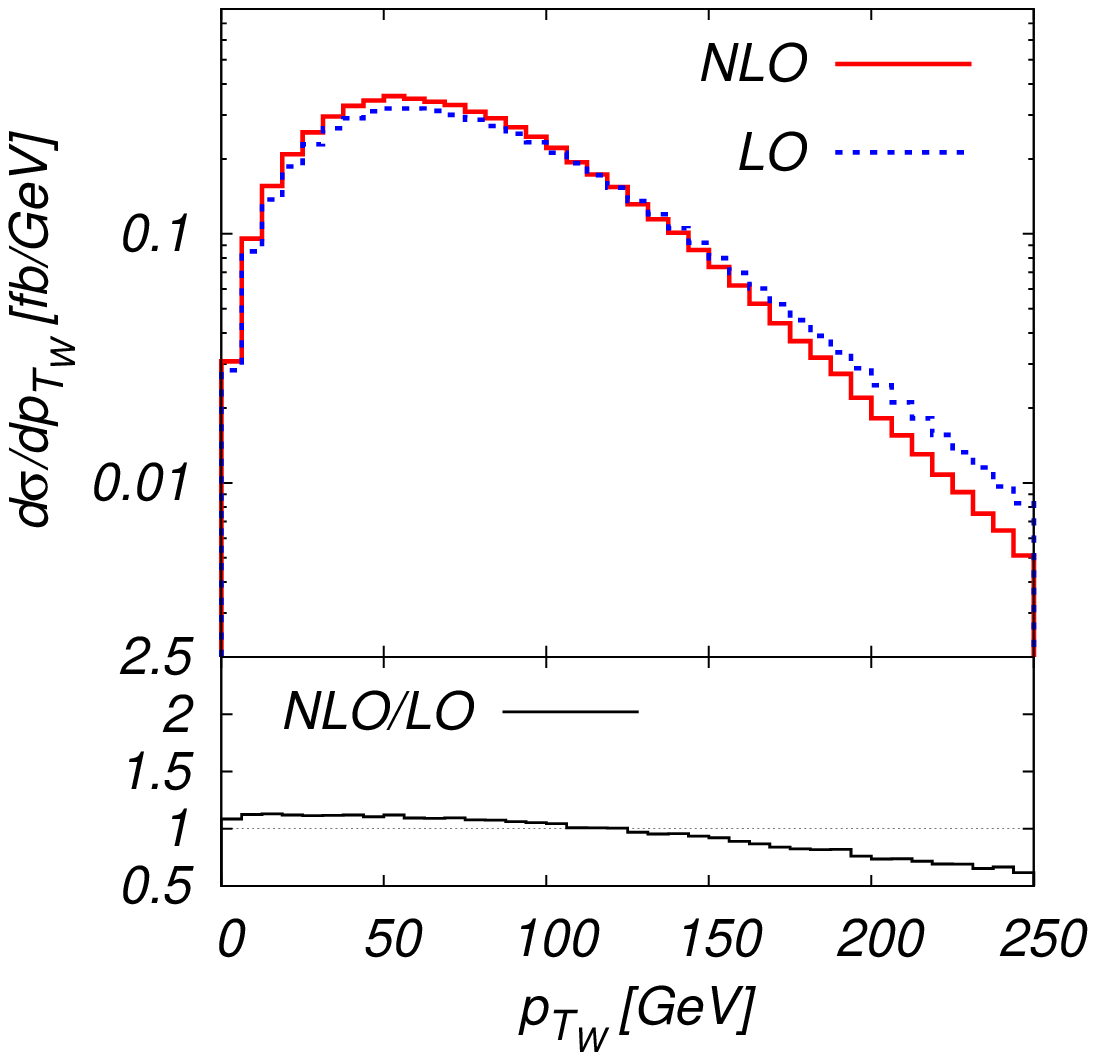}
\includegraphics[width=0.49\textwidth]{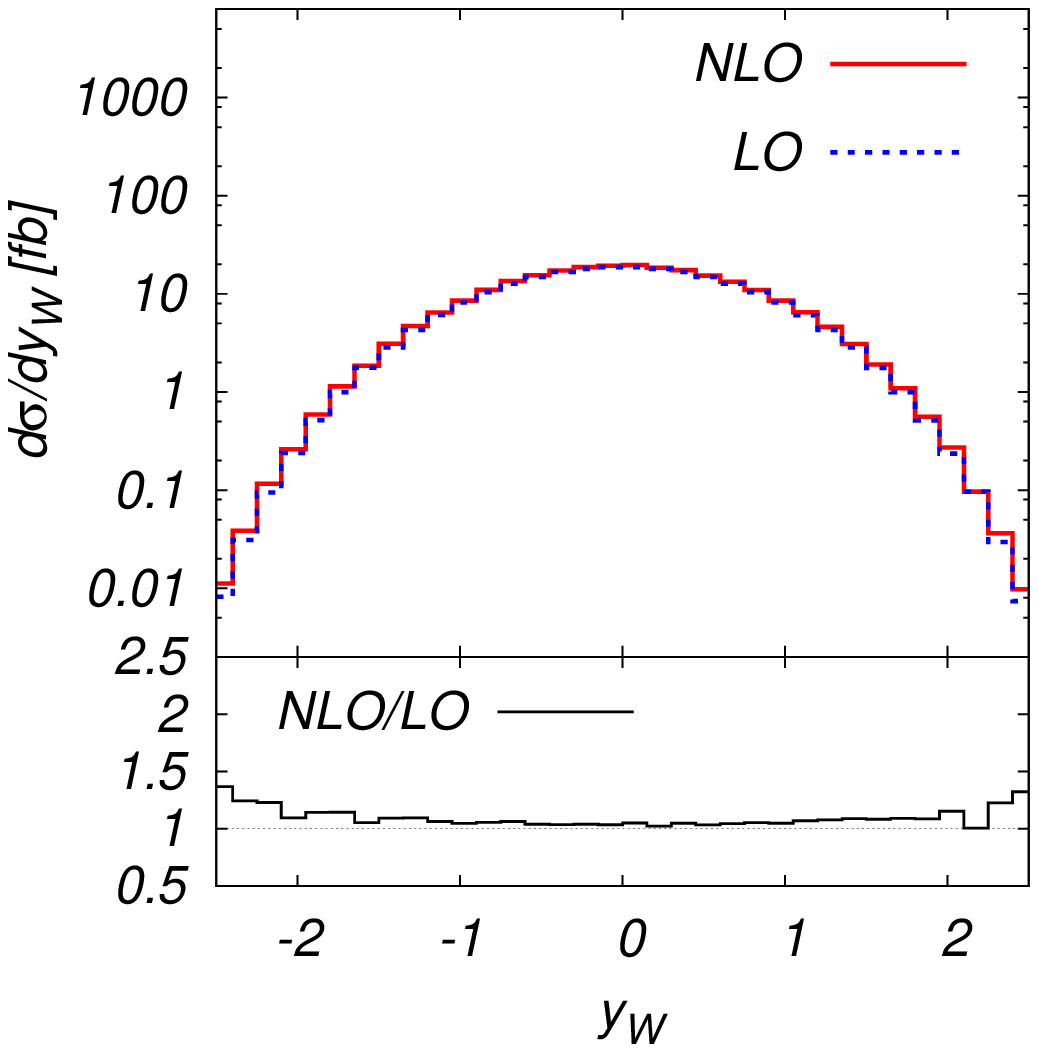}
\end{center}
\caption{\it \label{fig:W-tev} Differential  cross section
  distributions as a function of the averaged transverse momentum
  $p_{T_{W}}$  of the $W^{\pm}$ bosons and  averaged  rapidity $y_{W}$
  of the $W^{\pm}$ bosons for the $p\bar{p}\rightarrow
  e^{+}\nu_{e}\mu^{-}\bar{\nu}_{\mu}b\bar{b} ~ + X$ process at the
  TeVatron  run II.  The blue dashed curve corresponds to the leading
  order, whereas the red solid one to the next-to-leading order
  result. The lower panels display the differential K factor.  }
\end{figure}
\begin{figure}[th]
\begin{center}
\includegraphics[width=0.49\textwidth]{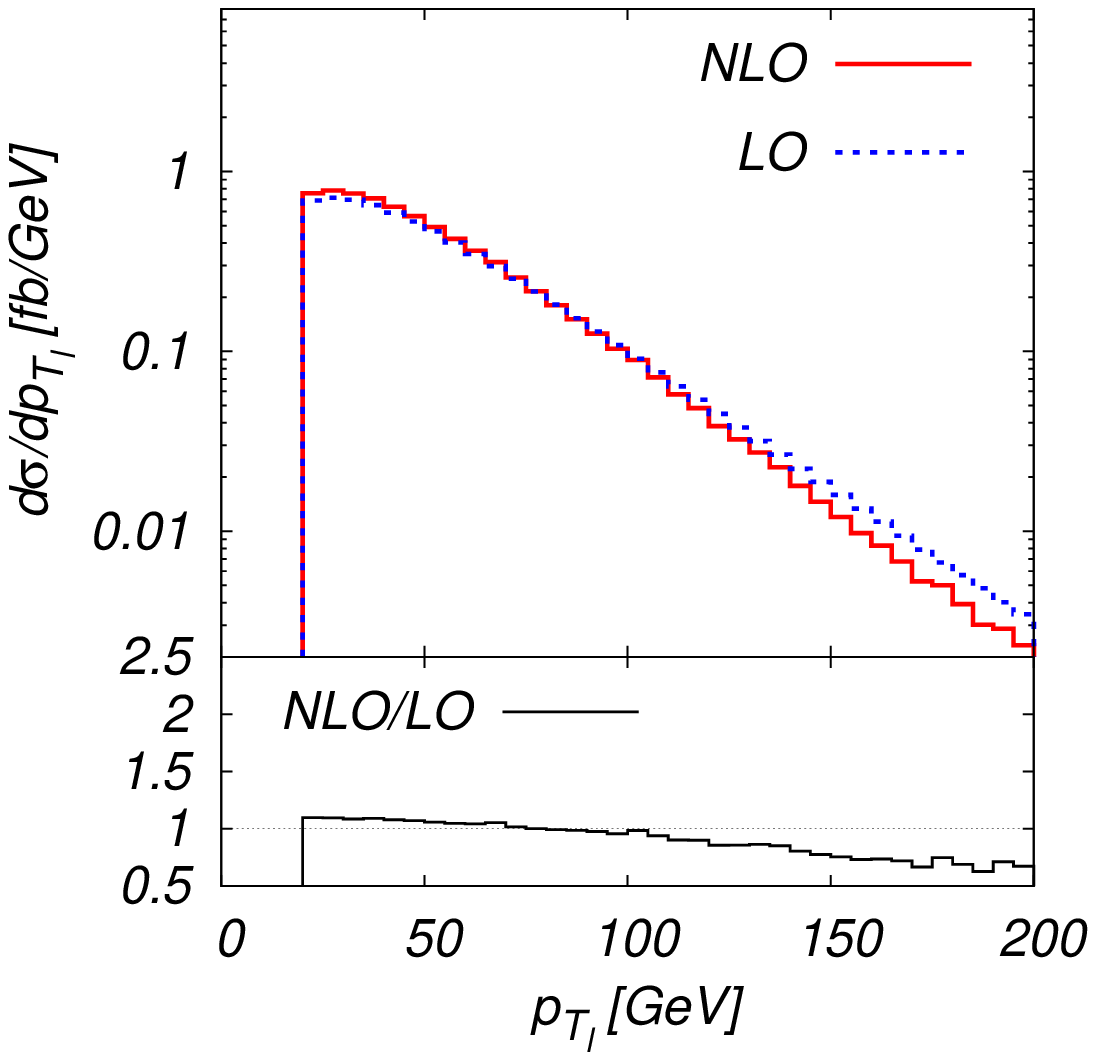}
\includegraphics[width=0.49\textwidth]{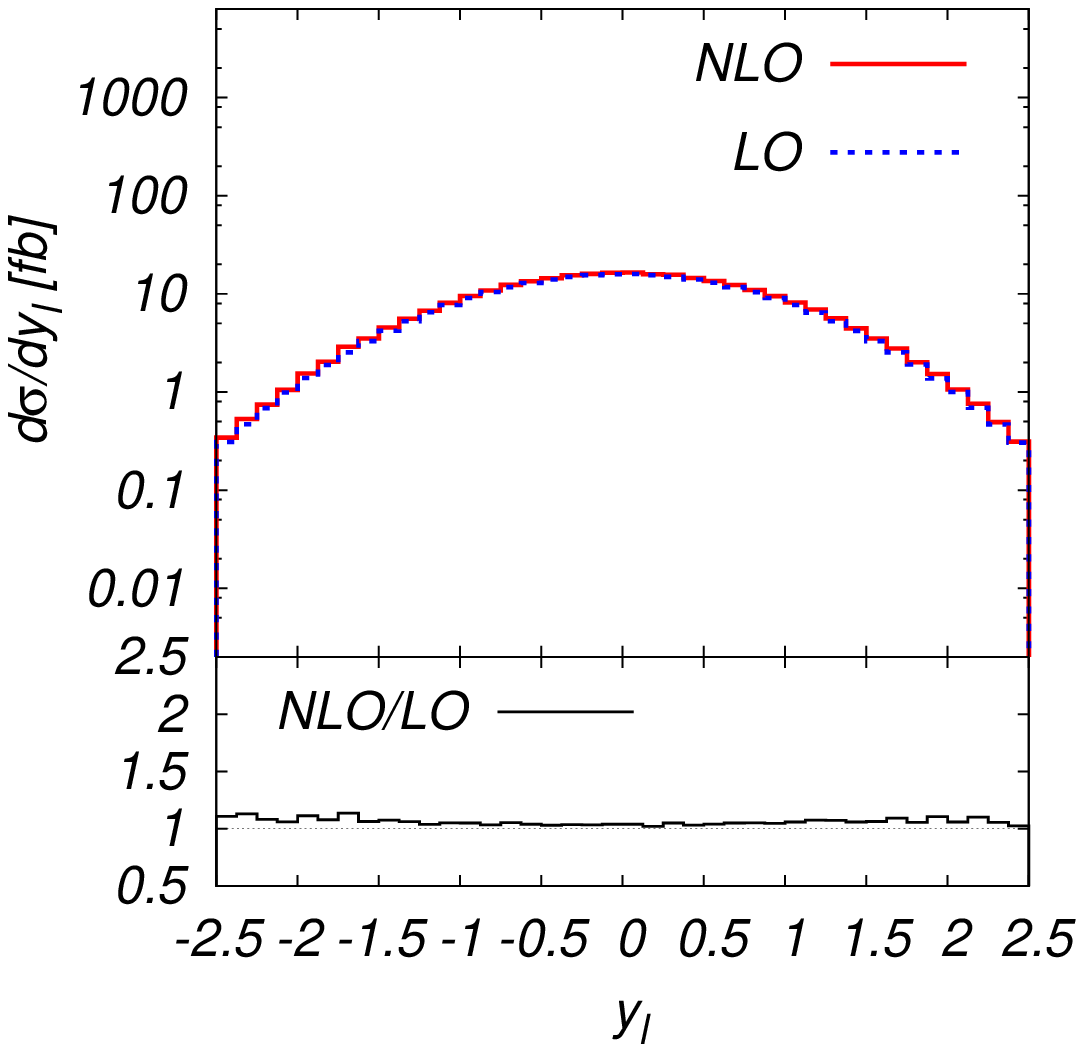}
\includegraphics[width=0.49\textwidth]{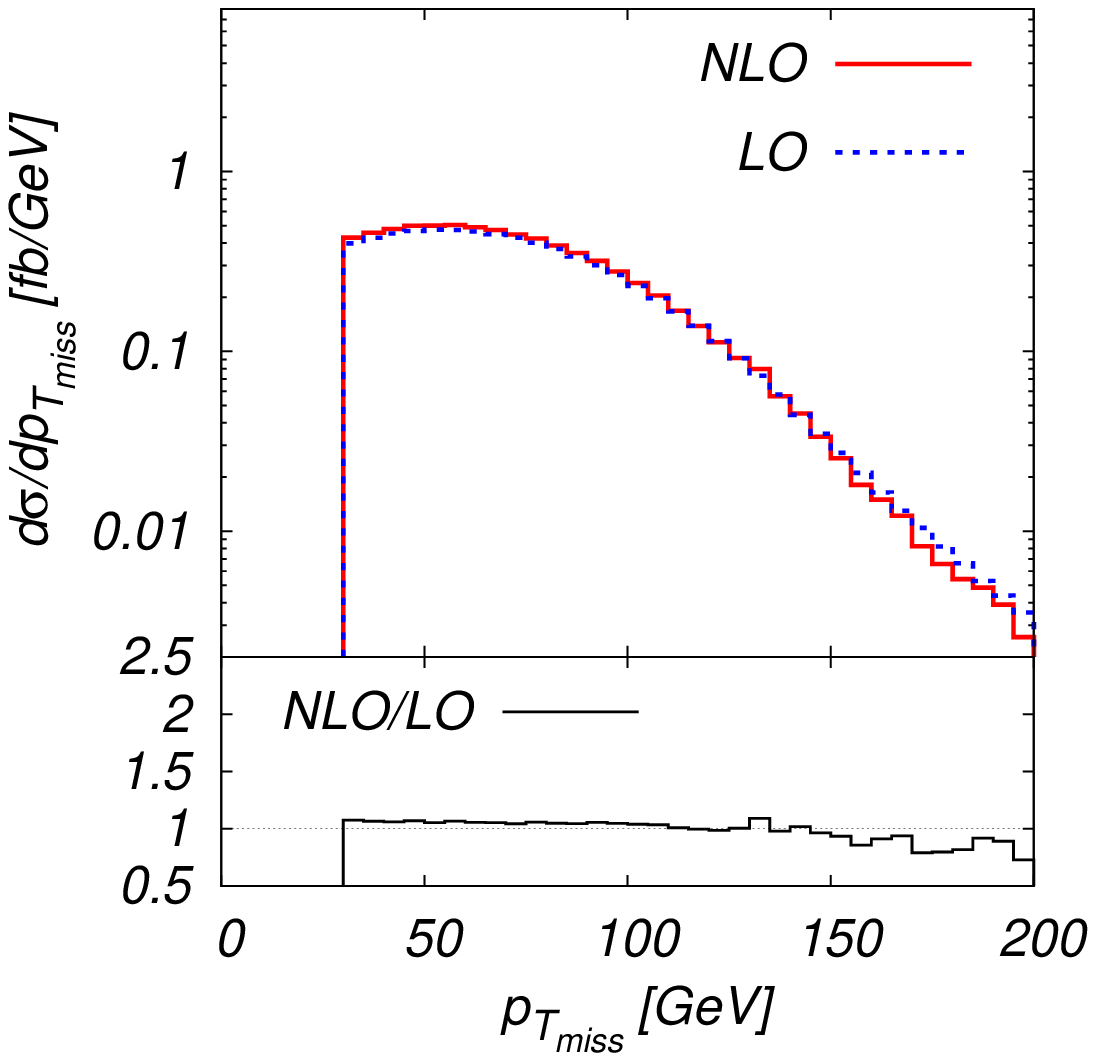}
\includegraphics[width=0.49\textwidth]{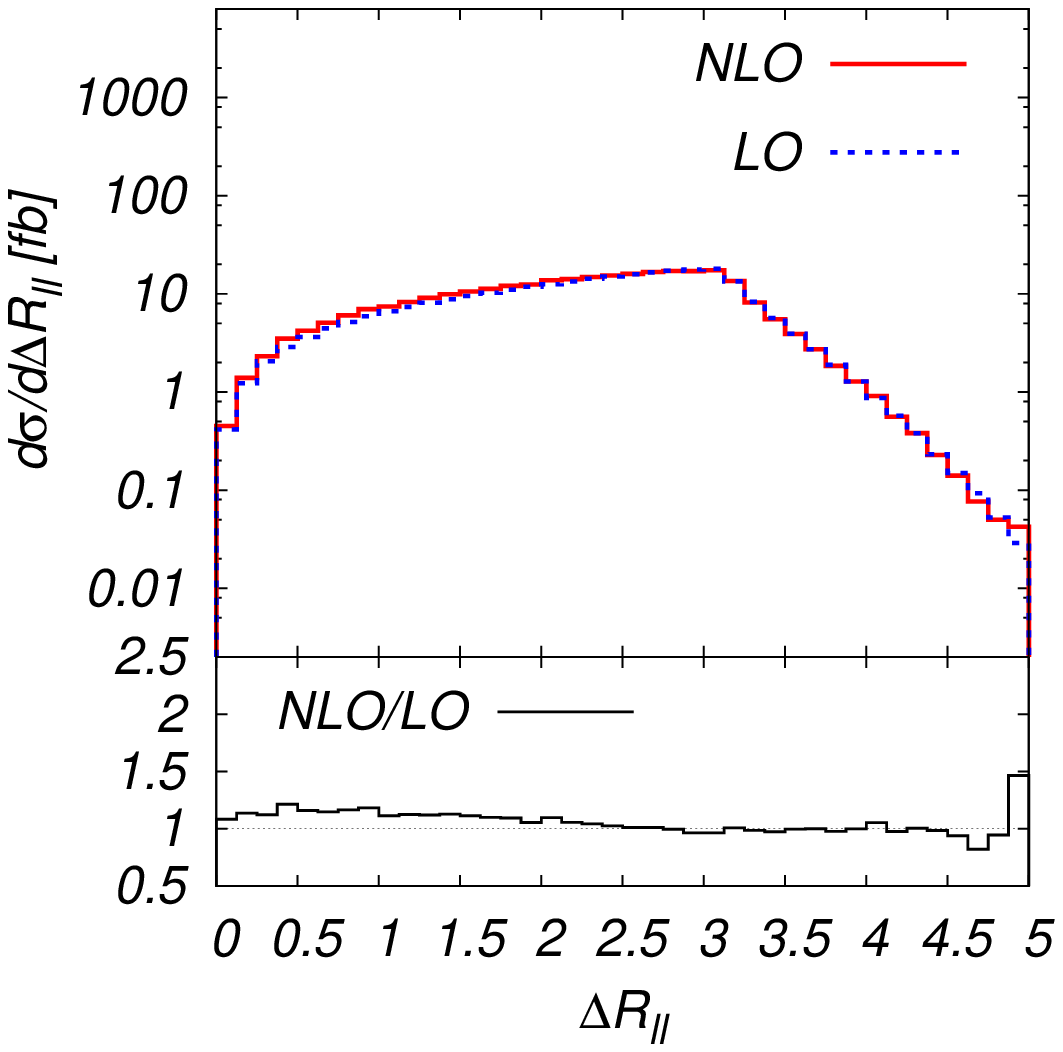}
\end{center}
\caption{\it \label{fig:lepton-tev}  Differential  cross section
  distributions as a function of the averaged transverse momentum
  $p_{T_{\ell}}$  of the  charged leptons,  averaged  rapidity
  $y_{\ell}$ of the  charged leptons, $p_{T_{miss}}$ and $\Delta
  R_{\ell\ell}$  for the $p\bar{p}\rightarrow
  e^{+}\nu_{e}\mu^{-}\bar{\nu}_{\mu}b\bar{b} ~ + X$ process at the
  TeVatron run II.  The blue dashed curve corresponds to the leading
  order, whereas the red solid one to the next-to-leading order
  result. The lower panels display  the differential K factor. }
\end{figure}
\begin{figure}[th]
\begin{center}
\includegraphics[width=0.49\textwidth]{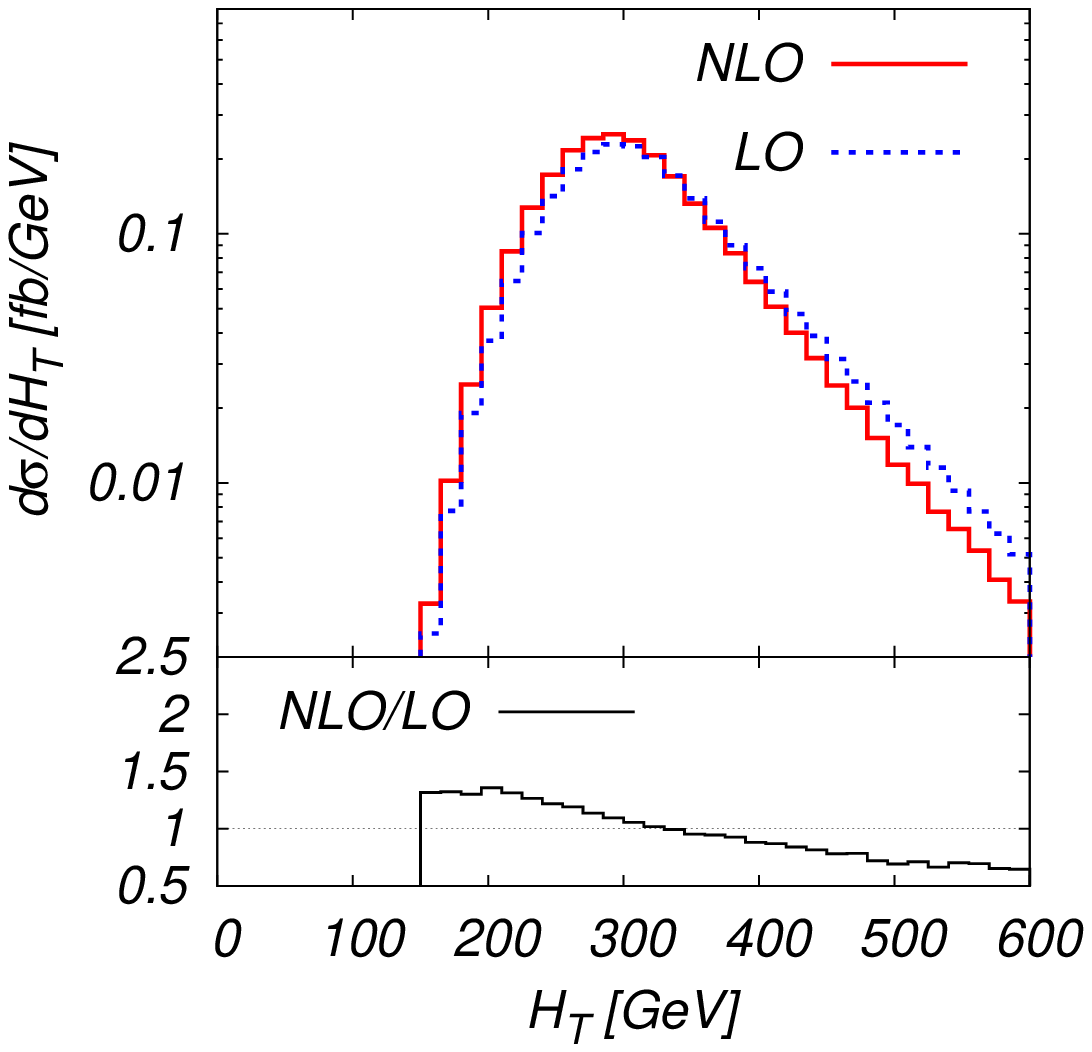}
\end{center}
\caption{\it \label{fig:hT-tev}  Differential  cross section
  distribution as a function of the total transverse energy, $H_T$,
  for the  $p\bar{p}\rightarrow
  e^{+}\nu_{e}\mu^{-}\bar{\nu}_{\mu}b\bar{b} ~ + X$ process at the
  TeVatron run II.  The blue dashed curve corresponds to the leading
  order, whereas the red solid one to the next-to-leading order
  result. The lower panels display the differential K factor.}
\end{figure}

In the following we would like to estimate the size of the
non-factorizable corrections for our inclusive setup. To achieve this
the full result has been compared with the result in the  NWA. The
latter has been  obtained by rescaling the coupling of the top quark
to the $W$ boson and the b quark by several large factors to mimic the
limit $\Gamma_t \to 0 $ when the scattering cross section factorizes
into on-shell production and decay.  Our findings are depicted in
Figure \ref{fig:rescaling-tev} where the dependence of the total NLO
cross section together with its individual contributions, real
emission part and LO plus virtual corrections, are  shown. The
behavior is compatible with a logarithmic dependence on   $\Gamma_t$,
which cancels between real and virtual corrections.  For inclusive
production, advancing from NWA to the full result changes the cross
section no more than  $+1\%$  which is consistent with the uncertainty
of the NWA {\it i.e.} of order  ${\cal {O}}(\Gamma_t/m_t)$.

Comparing our NLO integrated cross section with the value
$\sigma_{\rm NLO} = 36.47$ fb presented  in
Ref.~\cite{Melnikov:2009dn}, we observe a $2\%$ discrepancy, which can
easily be explained by two effects.  First of all, in
\cite{Melnikov:2009dn} NLO QCD corrections have been calculated
employing an on-shell approximation for the top quarks and the $W$
bosons. The former  approximation can introduce a difference of the
order of  ${\cal{O}}(\Gamma_t/m_t) \sim 1 \%$ while the latter of the
order of ${\cal{O}}(\Gamma_W/m_W) \sim 3 \%$.  As a second effect,
there are small differences  between individual setups, in {\it e.g.}
the value of $\Gamma_{t}$, $m_t$, $p_{T_{miss}}$ and $\Delta
R_{j\ell}$.

We have also compared our results with those generated with
\textsc{Mcfm}. We have been able to use the same cuts and input
parameters, but there is an essential difference as far as the
construction of the cross section is concerned. Indeed, \textsc{Mcfm}
includes corrections to the production of on-shell top quarks only,
whereas decays are included at leading order. Moreover, $W$ bosons are
also treated in the narrow width approximation. In the end,
\textsc{Mcfm} gives the following results $\sigma_{\rm LO}=(36.494 \pm
0.050)$ fb and $\sigma_{\rm NLO} = (39.622 \pm 0.065)$ fb, which are
different from ours by $4.5\%$ at LO and  by $11\%$ at NLO. 
Although we have not quantified the impact of different approaches
used, related to the  top quark and the W-boson finite width, as well
as the NLO corrections to the decay of the top quarks, the overall
comparison seems reasonable and compatible with estimates based  on
the order of magnitude for these effects, ${\cal {O}}(\Gamma_t/m_t)$,
${\cal {O}}(\Gamma_W/m_W)$ and ${\cal {O}}(\alpha_s)$.  
A more detailed study would be
necessary in order to establish the relevance of these differences for
the experimental analysis, which goes beyond the purpose of the
present publication.
 
In a next step we recalculate the top quark forward-backward asymmetry
for the TeVatron from the top rapidity distribution.  We show our
results  for the LO and NLO  inclusive calculations. At LO, $t\bar{t}$
production is totally  charge-conjugation symmetric for both
production mechanisms (quark and gluon  fusion). As a consequence, the
angular distributions of the $t$ and $\bar{t}$  are symmetric with
respect to the beam axis for $p\bar{p}$ collisions. However,  at
higher orders  in $\alpha_s$, this is not longer true.  Not all
processes involving additional partons are symmetric under charge
conjugation with respect to the incoming  parton and anti-parton
beams. As was pointed out in
Ref.\cite{Halzen:1987xd,Kuhn:1998jr,Kuhn:1998kw} the process
$gg\rightarrow t\bar{t}g$ is, but the processes $q\bar{q} \rightarrow
t\bar{t}g$  and $qg \rightarrow t\bar{t}q$ are not. Processes
involving initial state  valence quarks will therefore exhibit a
charge asymmetry.  This is  caused by interference between initial and
final state gluon emission on the one side and by interference between
color singlet 4-point virtual corrections and the Born term for the
$q\bar{q}$ process \cite{Nason:1987xz,Nason:1989zy} on the other.
Because $t\bar{t}$ production at the  TeVatron is dominated at the $95
\%$ level by $q\bar{q}$ annihilation, as was mentioned earlier in the
paper, we can expect the $q\bar{q}$ subprocess asymmetry to be visible
in the total sample. The integrated charge asymmetry is defined
through 
\begin{equation}
A=\frac{\int_{y_t>0}N_{t}(y)-\int_{y_{\bar t}>0}N_{\bar{t}}(y)}{\int_{y_t>0}
  N_{t}(y)+\int_{y_{\bar{t}}>0}N_{\bar{t}}(y)} \, ,
\end{equation}
where $y_t$ ($y_{\bar{t}}$) is the rapidity of the top (anti-top) quark in the
laboratory frame and $N_{t}(y)=d\sigma_{t\bar{t}}/dy_t$, 
$N_{\bar{t}}(y)=d\sigma_{t\bar{t}}/dy_{\bar{t}}$. Due to the
${\cal{CP}}$ invariance of QCD the rapidity  distributions of top and
anti-top are mirror images of each other, {\it i.e.}
$N_{\bar{t}}(y) = N_{t}(-y)$, and integrated charge
asymmetry is equal to the integrated forward-backward  asymmetry of
the top quark defined as 
\begin{equation}
A^{t}_{FB}=\frac{\int_{y>0}N_{t}(y)-\int_{y<0}N_{t}(y)}{\int_{y>0}
  N_{t}(y)+\int_{y<0}N_{t}(y)}\, .
\end{equation}
Moreover, $A^{\bar{t}}_{FB}=- A^{t}_{FB}$.  

As can be seen in the
upper-left part of the  Figure \ref{fig:asymmetry} the LO $t\bar{t}$ inclusive
cross section is  symmetric around $y_t=0$ (green dashed curve).  The NLO
inclusive result for the top/anti-top quark is,  on the other hand, shifted to
larger $y_t$ for the top quark (solid red curve)  and smaller $y_t$ for the 
anti-top quark (dotted blue curve). This corresponds  to a positive integrated
forward-backward asymmetry of the order of 
\begin{equation}
A^{t}_{FB} = 0.051 \pm 0.0013\, , 
\end{equation}
which tells us that top quarks are preferentially emitted in the direction of
the incoming protons.

Next-to-leading order contributions to 
the forward-backward asymmetry have already  been calculated 
in the on-shell $t\bar{t}$ production \cite{Antunano:2007da} and 
amount to $A^{t}_{FB}=0.051 \pm 0.006$. The CDF
measurement based on $5.3$ fb$^{-1}$ integrated luminosity  in the
semi-leptonic channel  yields $A^{t}_{FB}=0.150 \pm 0.050^{stat.} \pm
0.024^{syst.}$ \cite{CDF},  while the  D\O{} measurement of this asymmetry
yields $A^{t}_{FB}=0.08 \pm 0.04^{stat.} \pm 0.01^{syst.}$  based on
$4.3$ fb$^{-1}$ integrated luminosity  \cite{D0}.  The
uncertainties of these results are still very large and statistically
dominated. 

In the same manner we can calculate  the  integrated forward-backward
asymmetry for the top decay products,  namely the $b$-jet and the
positively  charged lepton.  Our results can be summarized as follows:
\begin{equation}
A^b_{FB}=  0.033  \pm 0.0013\,  , ~~~~~ A^{\ell^{+}}_{FB}= 0.034  \pm
0.0013\, ,
\end{equation}
where 
\begin{equation}
A^{b}_{FB}=\frac{\int_{y_{b}>0}N_{b}(y)-\int_{y_{b}<0}N_{b}(y)}{\int_{y_{b}>0}
  N_{b}(y)+\int_{y_{b}<0}N_{b}(y)}\,  , ~~~~~
A^{\ell^{+}}_{FB}=\frac{\int_{y_{\ell^{+}}>0}N_{\ell^{+}}(y)-\int_{y_{\ell^{+}}<0}
  N_{\ell^{+}}(y)}{\int_{y_{\ell^{+}}>0}
  N_{\ell^{+}}(y)+\int_{y_{\ell^{+}}<0}N_{\ell^{+}}(y)}
\end{equation}
and $y_\ell$ and $y_b$ are the rapidity of the charged lepton and the
$b$-jet respectively and
$N_{\ell^{+}}(y)=d\sigma_{t\bar{t}}/dy_{\ell^{+}}$,
$N_b(y)=d\sigma_{t\bar{t}}/dy_b$. 
In case of $A^{\ell^{+}}_{FB}$ we agree with Ref.\cite{Bernreuther:2010ny} where
$A^{\ell^{+}}_{FB}=0.033$ has been quoted. The integrated forward-backward
asymmetries  of the charged lepton and the $b$-jet  have the same sign
as $A^{t}_{FB}$ but are smaller in magnitude. Let us stress at this
point, that the $b$-jet integrated forward-backward asymmetry is a
rather theoretical observable even though it can in principle be
measured once  the $b$-jet is distinguished experimentally from the
anti-$b$-jet through {\it e.g.} the charge of the associated lepton
flying in the same direction. However, it is extremely difficult to
determine the charge of the $b$-jet and this measurement will heavily
depend on the $b$-jet tagging efficiency. The $b$-jet and charged
lepton differential distributions  in rapidity are also presented in
Figure \ref{fig:asymmetry}. 

While the size of the corrections to the total cross section is certainly 
interesting, it is crucial to study the corrections to distributions. 
In the following, the NLO QCD corrections to the differential distributions 
for the dileptonic channel with full off-shell effects are presented.

In Figure \ref{fig:top-tev} we start with the most important
observable, namely, the differential distribution of the $t\bar{t}$
invariant mass, $m_{t\bar{t}}$.  Figure \ref{fig:top-tev} depicts also
the rapidity, $y_{t\bar{t}}$,  of the top-anti-top system as  well as
the averaged transverse momentum, $p_{T_{t}}$, and  the averaged
rapidity $y_{t}$ of  the top and anti-top. The blue dashed curve
corresponds to the leading order, whereas the red solid one to the
next-to-leading order result.  The histograms can also be turned into
dynamical K-factors, which we display in the lower panels. The small
size of the corrections to the total cross section is reflected only
in the angular distributions, where we can see positive corrections of
the order of  $5\%-10\%$. Both distributions of $m_{t\bar{t}}$ and
$p_{T_t}$ get sizeable negative corrections for large values of these
observables.  For the $m_{t\bar{t}}$ distribution, corrections reach
$-30\%$ which has to be compared  with positive $+25\%$   corrections
close to the $t\bar{t}$ threshold.  The $p_{T_{t}}$ distribution is
corrected down to $-40\%$ at the tails  and  $+20\%$ for small  values
of $p_{T_{t}}$. Overall, this leads  to a distortion of the
differential distributions up to $55\%-60\%$.  Given that top-quark
pair production at high scale is an ideal tool to search for various
models of physics beyond the Standard Model with new gauge bosons like
{\it e.g.} $Z^{\prime}$,  it is clear that a precise knowledge of  the
higher order corrections in this region  is of significant
importance.

In Figure \ref{fig:bottom-tev}, the $b$-jet kinematics is presented,
where differential  cross section distributions as a function of  the
averaged transverse momentum, $p_{T_{b}}$, and averaged rapidity,
$y_{b}$,  of the $b$- and anti-$b$-jet are presented together with
the $\Delta R_{b\bar{b}}$ separation. Both angular distributions,
$y_{b}$ and $\Delta R_{b\bar{b}}$, exhibit small positive corrections
$5\%-10\%$,  however, for the  $p_{T_b}$ distribution we observe large
and positive corrections of the order of $+30\%$ at the begin of the
spectrum and negative of the order of $-20\%$  around $200$ GeV.

A similar situation is observed for the $W^{\pm}$ boson kinematics
which is shown in Figure \ref{fig:W-tev}, where  the differential
cross section distributions as function of  the averaged transverse
momentum  $p_{T_{W}}$ of the $W^{\pm}$ bosons together with an
averaged  rapidity $y_{W}$ of the $W^{\pm}$ bosons are depicted. Yet
again, small positive corrections of $5\%-10\%$ are acquired for
angular distributions as well as   for low values of $p_{T_{W}}$,
while the tail of the $p_{T_{W}}$ differential distribution exhibits
negative corrections down to $-30\%$. 

Subsequently, in Figure \ref{fig:lepton-tev}, differential  cross
section distributions as function of the averaged transverse momentum
$p_{T_{\ell}}$ and   averaged  rapidity $y_{\ell}$ of the  charged
leptons together with $p_{T_{miss}}$ and the separation $\Delta
R_{\ell\ell}$ are shown. Also here, a distortion of the $p_{T_{\ell}}$
differential distribution up to $40\%$ is reached, while for
$p_{T_{miss}}$ up to  $15\%$. For the angular distributions,
moderate corrections up to $+10\%$ are obtained. 

And finally, in Figure \ref {fig:hT-tev}, the differential   cross
section distribution as function of the total transverse energy
defined as  
\begin{equation}
H_T = p_{T_{b}}+ p_{T_{\bar{b}}}+p_{T_{e^{+}}}+ p_{T_{\mu^{-}}}+
p_{T_{miss}}
\label{defHT}
\end{equation}
 is presented. In this case we observe a distortion of the
 differential distribution up to  $70\%-80\%$.

Overall, we can say that at the TeVatron, employing a fixed scale
$\mu=m_t$, the  NLO corrections to transverse  momentum distributions
are moderate.  However, they do not simply rescale the  LO shapes, but
induce distortions at the level of $15\%-80\%$, which redistribute
events from larger to smaller transverse momenta.  The same applies to
the invariant mass distribution of the $t\bar{t}$ pair.  As for
angular  distributions we observe positive and rather modest
corrections of the order  of $5\%-10\%$.

\subsection{Results for the LHC}
\label{sec:lhc}

\begin{figure}[th]
\begin{center}
\includegraphics[width=0.49\textwidth]{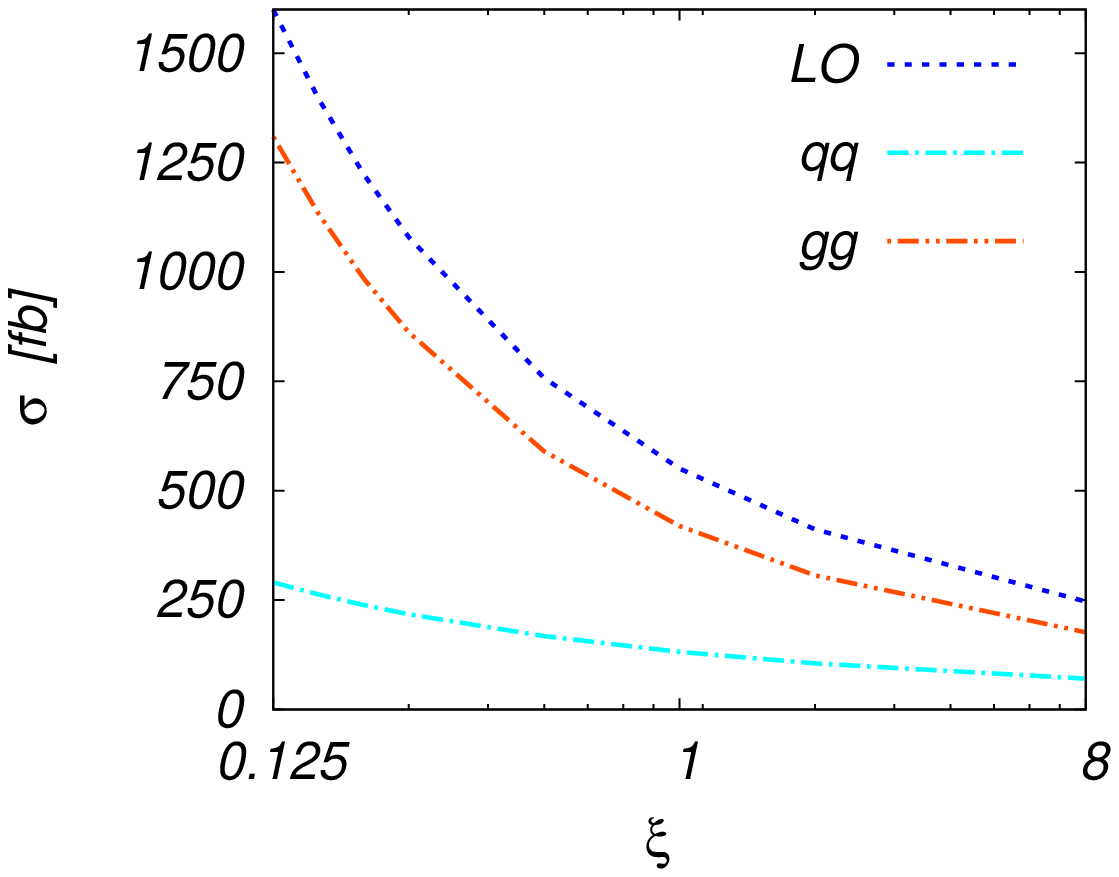}
\includegraphics[width=0.49\textwidth]{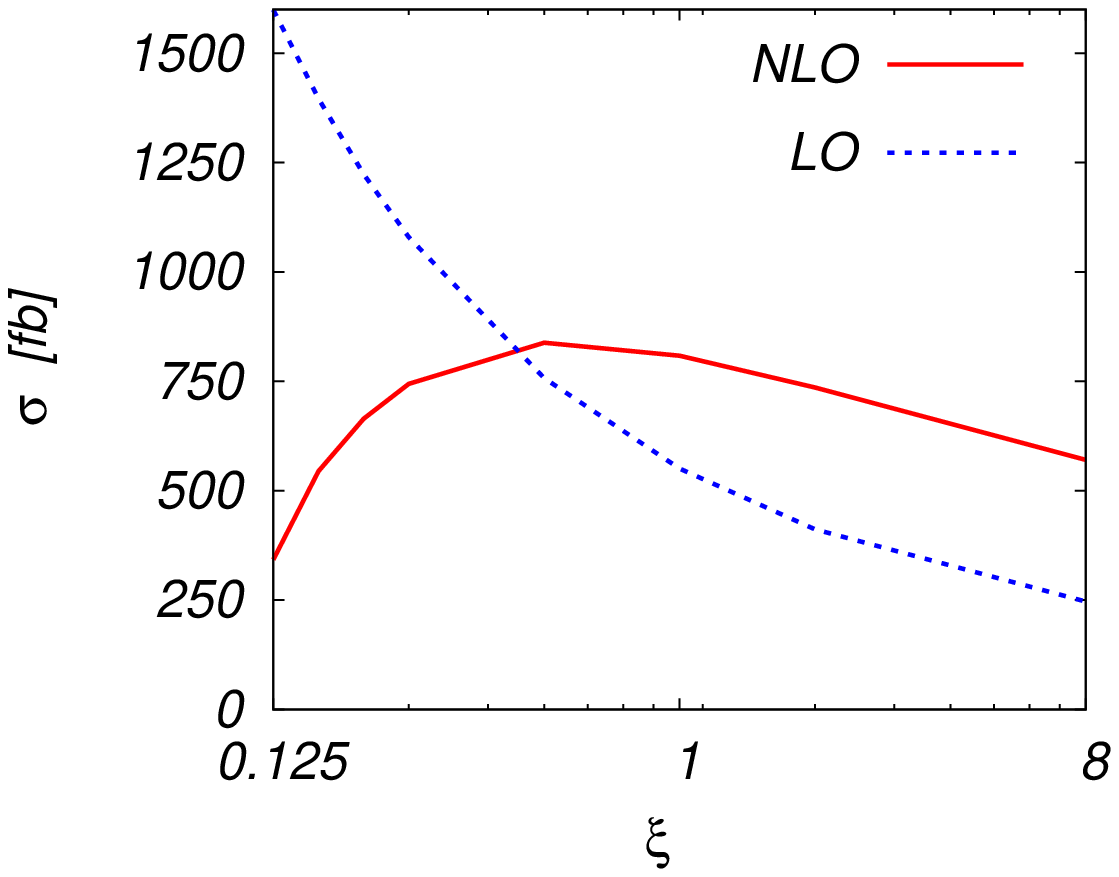}
\end{center}
\caption{\it \label{fig:scales-lhc} Scale dependence of the LO cross
  section with the individual contributions of the partonic channels
  (left panel) and  scale dependence of the LO and NLO cross sections
  (right panel)  for the  $pp\rightarrow
  e^{+}\nu_{e}\mu^{-}\bar{\nu}_{\mu}b\bar{b} ~+ X$ process at the LHC
  with $\sqrt{s} = 7$ TeV, where renormalization  and factorization
  scales are set to the common value  $\mu=\mu_R=\mu_F=\xi m_t$.}
\end{figure}

Table \ref{tab:lhc} shows the integrated cross sections at the LHC
with $\sqrt{s}= 7$ TeV, for two choices of the $\alpha_{max}$
parameter and for three different jet algorithms.  At the central
scale value, the full cross section receives  NLO QCD  corrections  of
the order of $47\%$.  Figure \ref{fig:scales-lhc} presents the
dependence of  the integrated LO cross section on the renormalization
and factorization  scales where $\mu=\mu_R=\mu_F=\xi m_t$. The
variation range from  $\mu=m_t/8$ to $\mu=8 m_t$. In contrast to the
TeVatron, the gg channel comprises about  $76\%$  of the LO $pp$ cross
section, followed by the $q\bar{q}$ channel with about $24\%$.  In the
right panel of Figure \ref{fig:scales-lhc}, the scale dependence of
the NLO cross section is shown  together with the LO one. Comparing
the LO and NLO predictions, we find again that the large scale
dependence of about $+37\%$ and $-25\%$ in the LO cross section is
considerably reduced, down to $+4\%$ and $-9\%$ when varying the scale
down and up by a factor 2, after including the NLO corrections.

\begin{table}[th]
\begin{center}
  \begin{tabular}{|c|c|c|c|}
    \hline && & \\ Algorithm & $\sigma_{\rm LO}$ [fb]      &
    $\sigma_{\rm NLO}^{\rm \alpha_{max}=1}$  [fb] &
    $\sigma_{\rm NLO}^{\rm \alpha_{max}=0.01}$   [fb] \\ && &
    \\ \hline {\it anti}-$k_T$  & 550.54 $\pm$ 0.18 & 808.46 $\pm$
    0.98 & 808.29 $\pm$ 1.04 \\ \hline  $k_T$ & 550.54 $\pm$ 0.18 &
    808.67 $\pm$ 0.97 & 808.86 $\pm$ 1.03 \\ \hline C/A & 550.54 $\pm$
    0.18 &  808.74 $\pm$ 0.97 & 808.28 $\pm$ 1.03  \\ \hline
  \end{tabular}
\end{center}
 \caption{\it \label{tab:lhc} Integrated cross section at LO and NLO
   for  $pp\rightarrow e^{+}\nu_{e}\mu^{-}\bar{\nu}_{\mu}b\bar{b} ~+
   X$  production at the LHC with $\sqrt{s}= 7  ~\textnormal{TeV}$,
   for three different jet algorithms,  the anti-$k_T$, $k_T$ and for
   the Cambridge/Aachen jet algorithm.  The two NLO results refer to
   different values of  the dipole phase space cutoff
   $\alpha_{max}$. The scale choice is  $\mu_R=\mu_f=m_{t}$.}
\end{table}
\begin{figure}[th]
\begin{center}
\includegraphics[width=0.49\textwidth]{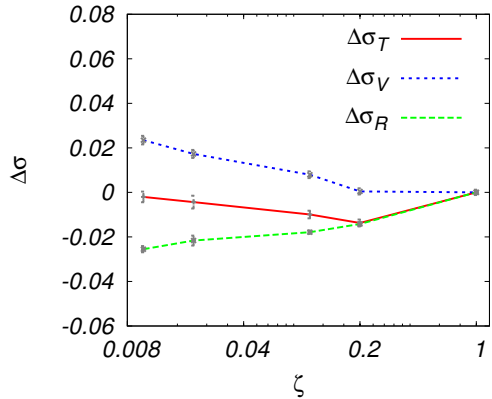}
\end{center}
\caption{\it \label{fig:rescaling-lhc} Dependence of the NLO   cross
  section, $\sigma_{\rm T}$, (red solid line) and the individual
  contributions, the real emission part, $\sigma_{\rm R}$,  (green
  dashed line)  and the LO plus virtual part, $\sigma_{\rm V}$, (blue
  dotted line),  on the rescaling parameter $\zeta$ defined as
  $\Gamma_{rescaled}=  \zeta \Gamma_t$ for the   $pp\rightarrow
  e^{+}\nu_{e}\mu^{-}\bar{\nu}_{\mu}b\bar{b} ~+ X$ process at the LHC
  with  $\sqrt{s}=7$ TeV.  $\Delta\sigma$ is defined as follows:
  $\Delta\sigma_i(\zeta)=(\sigma_{i}(\zeta)-\sigma_{i}
  (\zeta=1))/\sigma_{\rm T}(\zeta=1)$ with $i=V,R,T$.} 
\end{figure}
\begin{table}[th]
\begin{center}
  \begin{tabular}{|c|c|c|c|}
    \hline
      && & \\
      Algorithm & $\sigma_{\rm LO}$ [fb]      & 
     $\sigma_{\rm NLO}^{\rm \alpha_{max}=1}$  [fb] &   
      $\sigma_{\rm NLO}^{\rm \alpha_{max}=0.01}$   [fb] \\
      && & \\
    \hline
{\it anti}-$k_T$ 
& 1394.72  $\pm$  0.75 & 1993.3 $\pm$ 2.5  & 1993.9 $\pm$ 2.7 \\
    \hline 
  $k_T$ 
& 1394.72  $\pm$  0.75 & 1995.2 $\pm$ 2.5  & 1994.3 $\pm$ 2.7 \\
 \hline
 C/A 
& 1394.72  $\pm$  0.75 & 1995.0 $\pm$ 2.5  &  1994.3 $\pm$ 2.7\\
\hline
  \end{tabular}
\end{center}
 \caption{\it \label{tab:lhc2} Integrated cross section at LO and NLO
   for  $pp\rightarrow e^{+}\nu_{e}\mu^{-}\bar{\nu}_{\mu}b\bar{b} ~+
   X$  production at the LHC with $\sqrt{s}= 10  ~\textnormal{TeV}$,
   for three different jet algorithms,  the anti-$k_T$, $k_T$ and for
   the Cambridge/Aachen jet algorithm.  The two NLO results refer to
   different values of  the dipole phase space cutoff
   $\alpha_{max}$. The scale choice is  $\mu_R=\mu_f=m_{t}$.}
\end{table}
\begin{figure}[th]
\begin{center}
\includegraphics[width=0.49\textwidth]{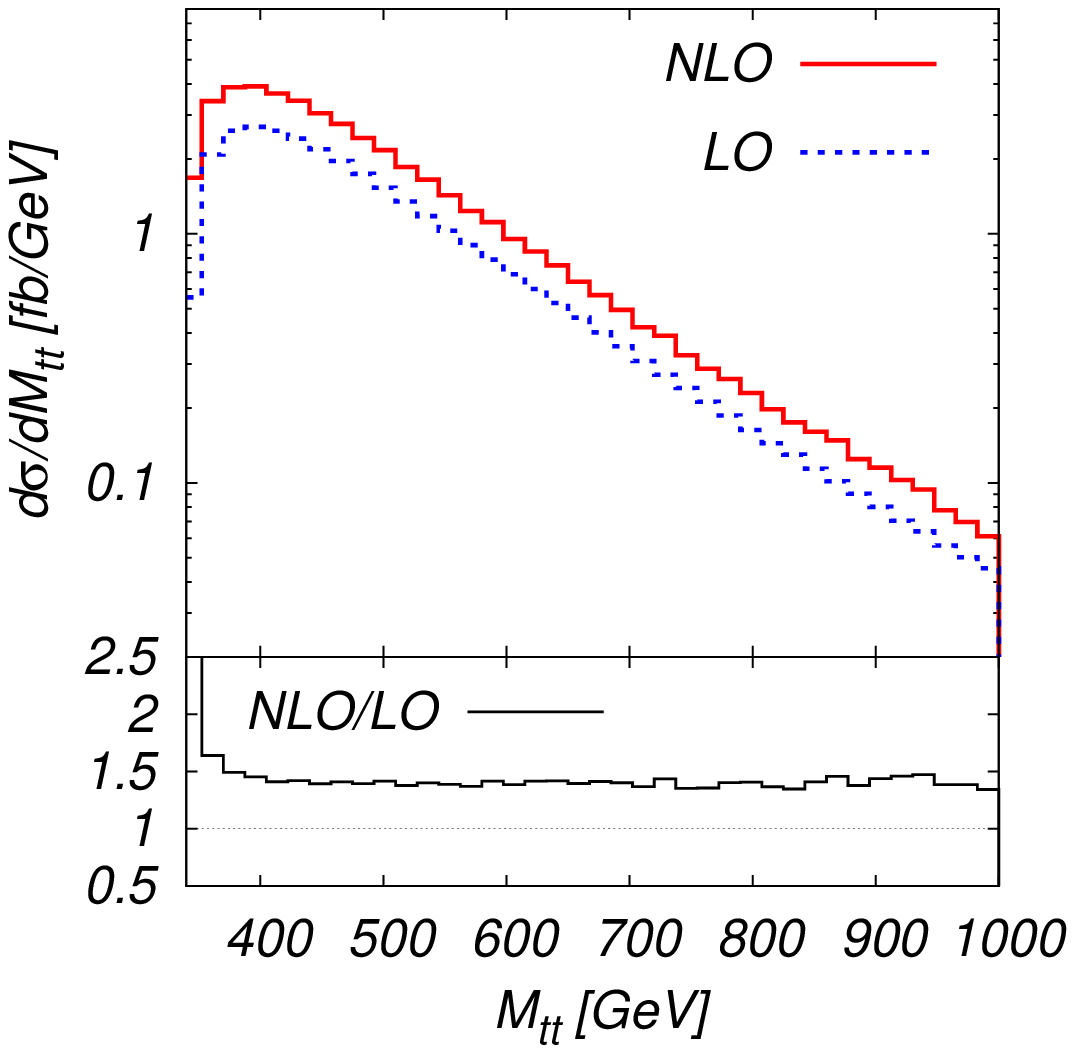}
\includegraphics[width=0.49\textwidth]{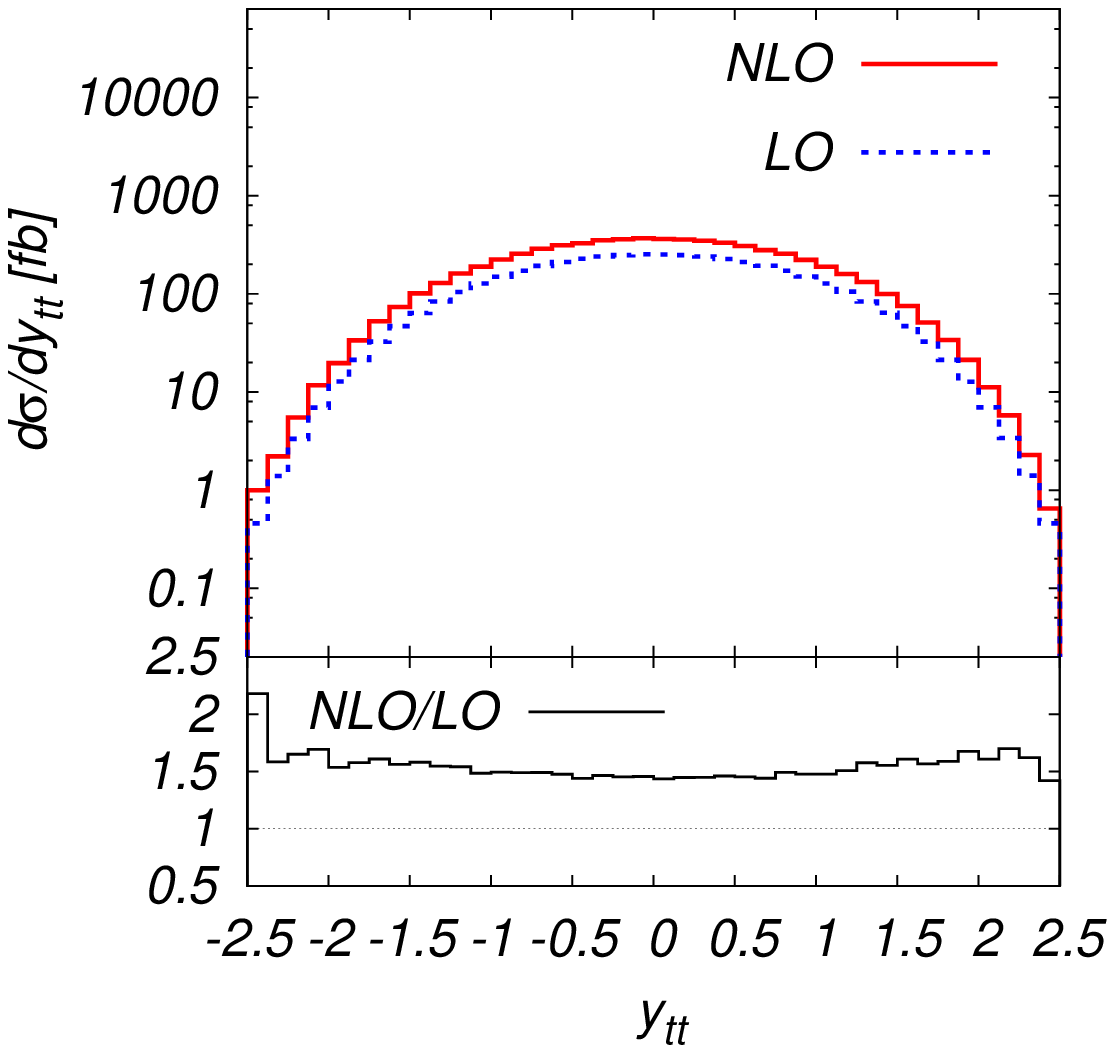}
\includegraphics[width=0.49\textwidth]{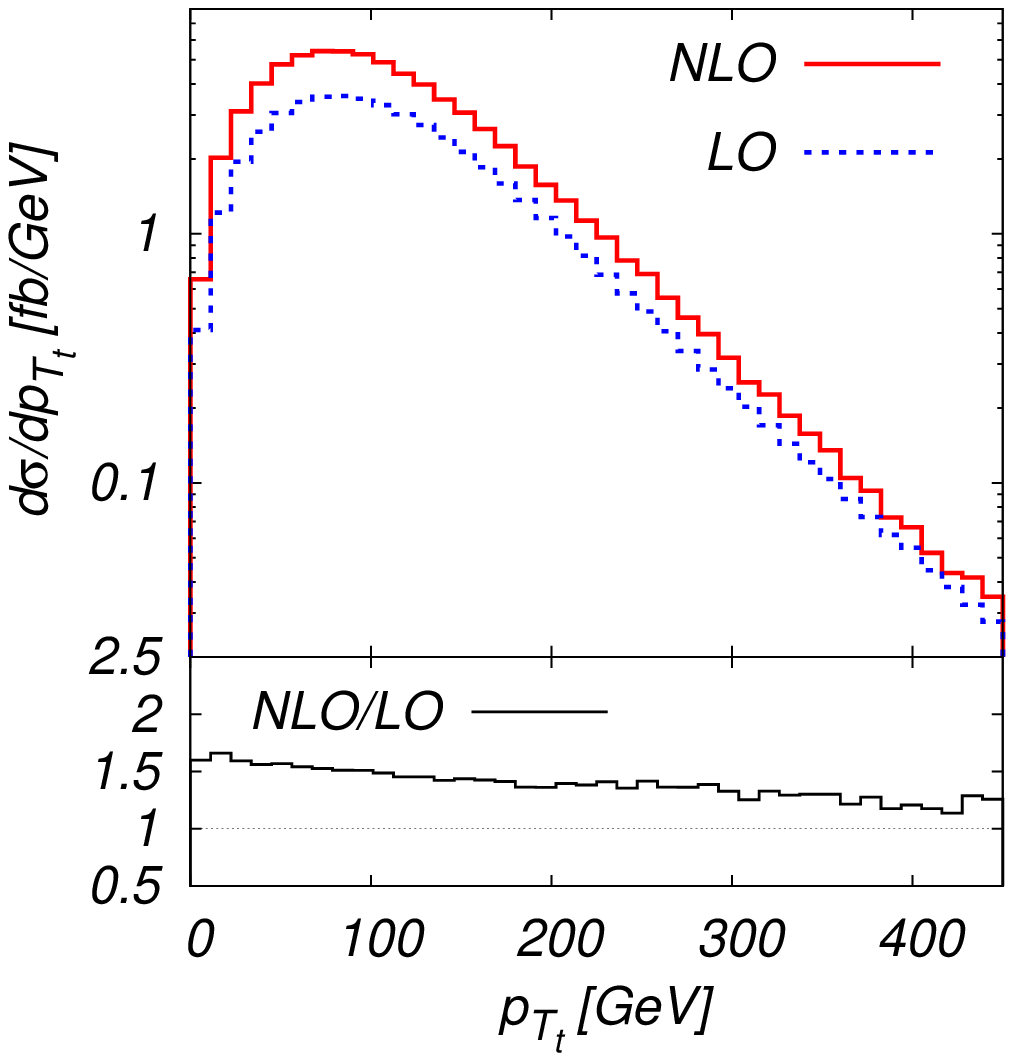}
\includegraphics[width=0.49\textwidth]{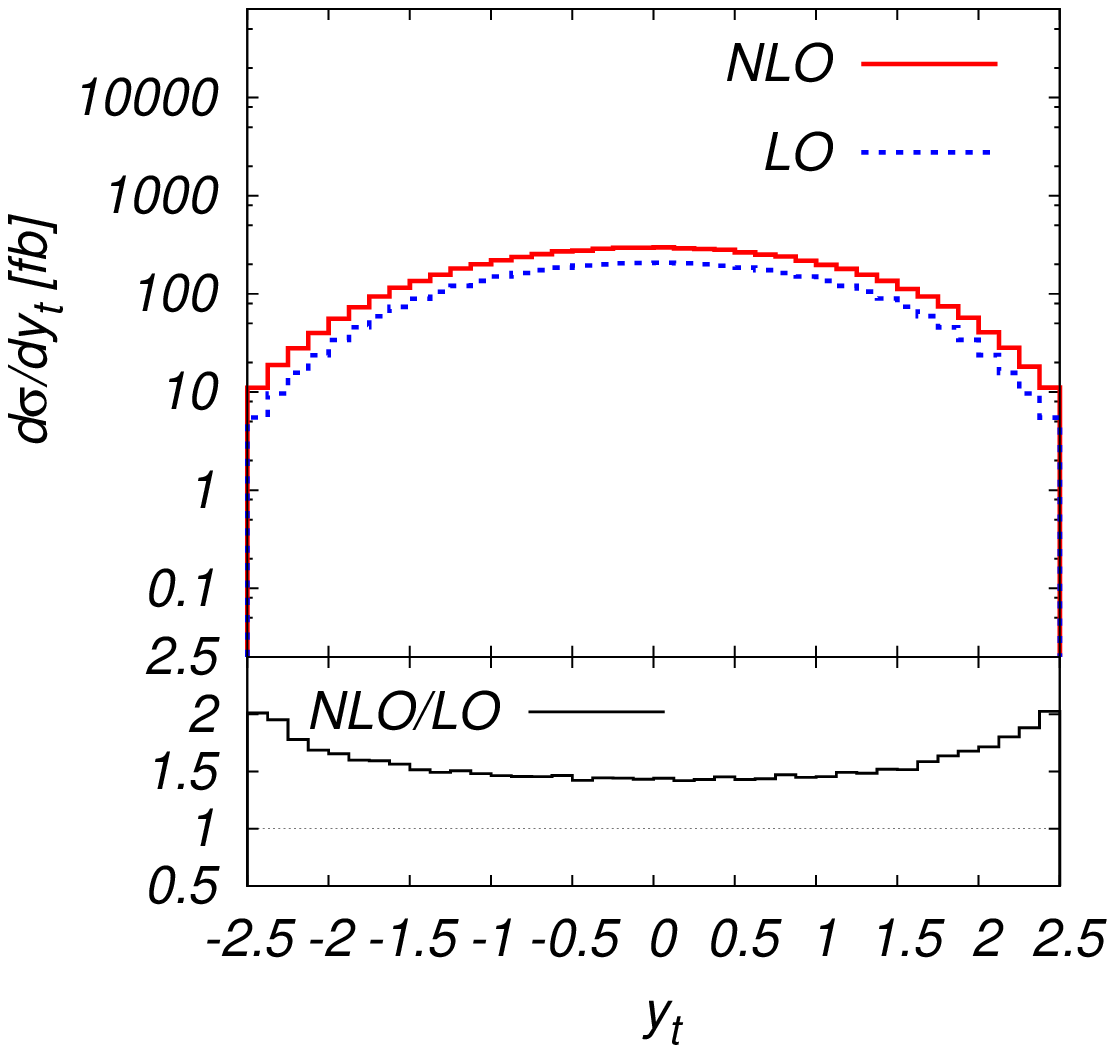}
\end{center}
\caption{\it \label{fig:top-lhc} Differential  cross section
  distributions as a function of the invariant mass $m_{t\bar{t}}$ of
  the top-anti-top pair, rapidity $y_{t\bar{t}}$ of the top-anti-top
  pair, averaged transverse momentum $p_{T_{t}}$  of the top and
  anti-top and  averaged  rapidity $y_{t}$ of  the top and anti-top
  for the $pp\rightarrow e^{+}\nu_{e}\mu^{-}\bar{\nu}_{\mu}b\bar{b} ~
  + X$ process at the LHC  with $\sqrt{s}= 7$ TeV.  The blue dashed
  curve corresponds to the leading order, whereas the red solid one to
  the next-to-leading order result. The lower panels display  the
  differential K factor.}
\end{figure}
\begin{figure}[th]
\begin{center}
\includegraphics[width=0.49\textwidth]{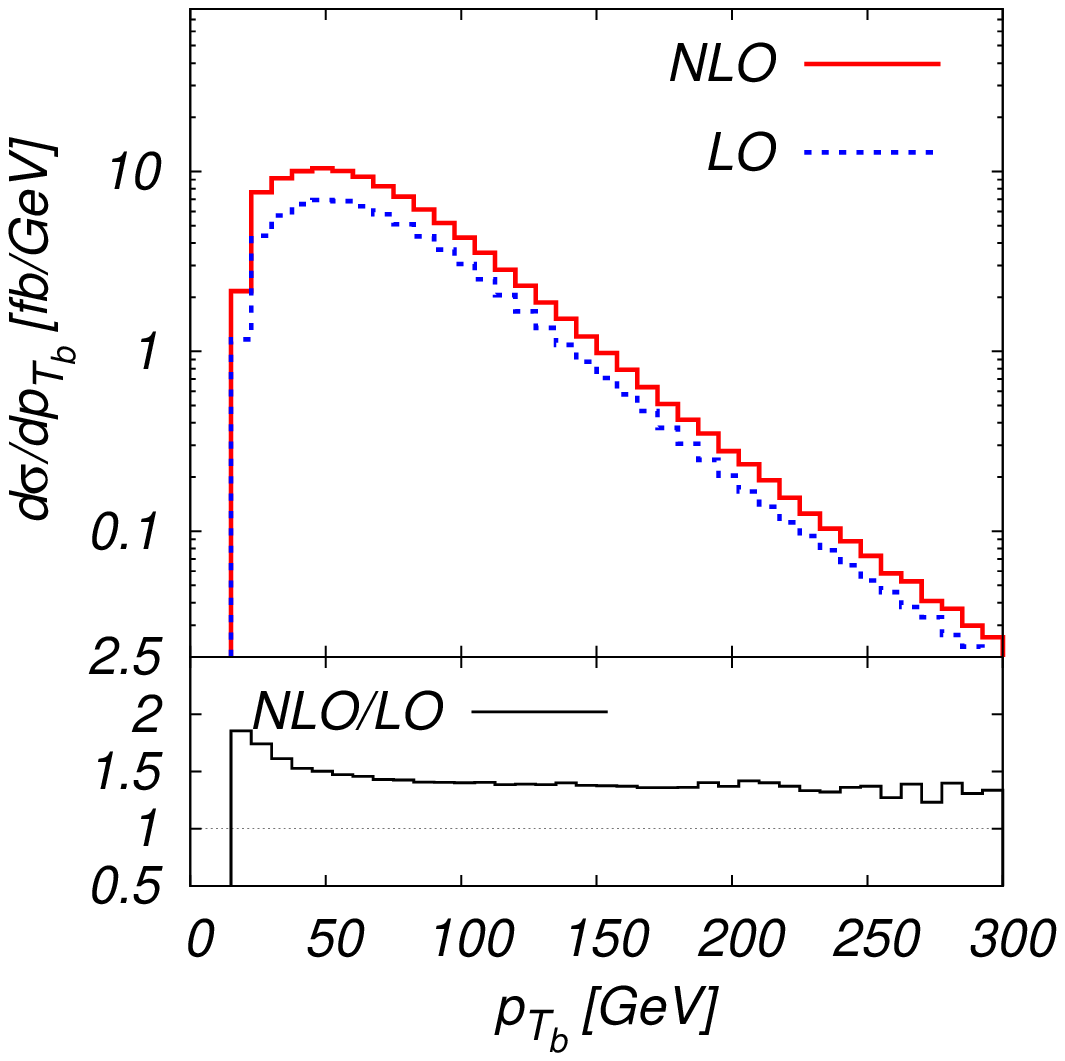}
\includegraphics[width=0.49\textwidth]{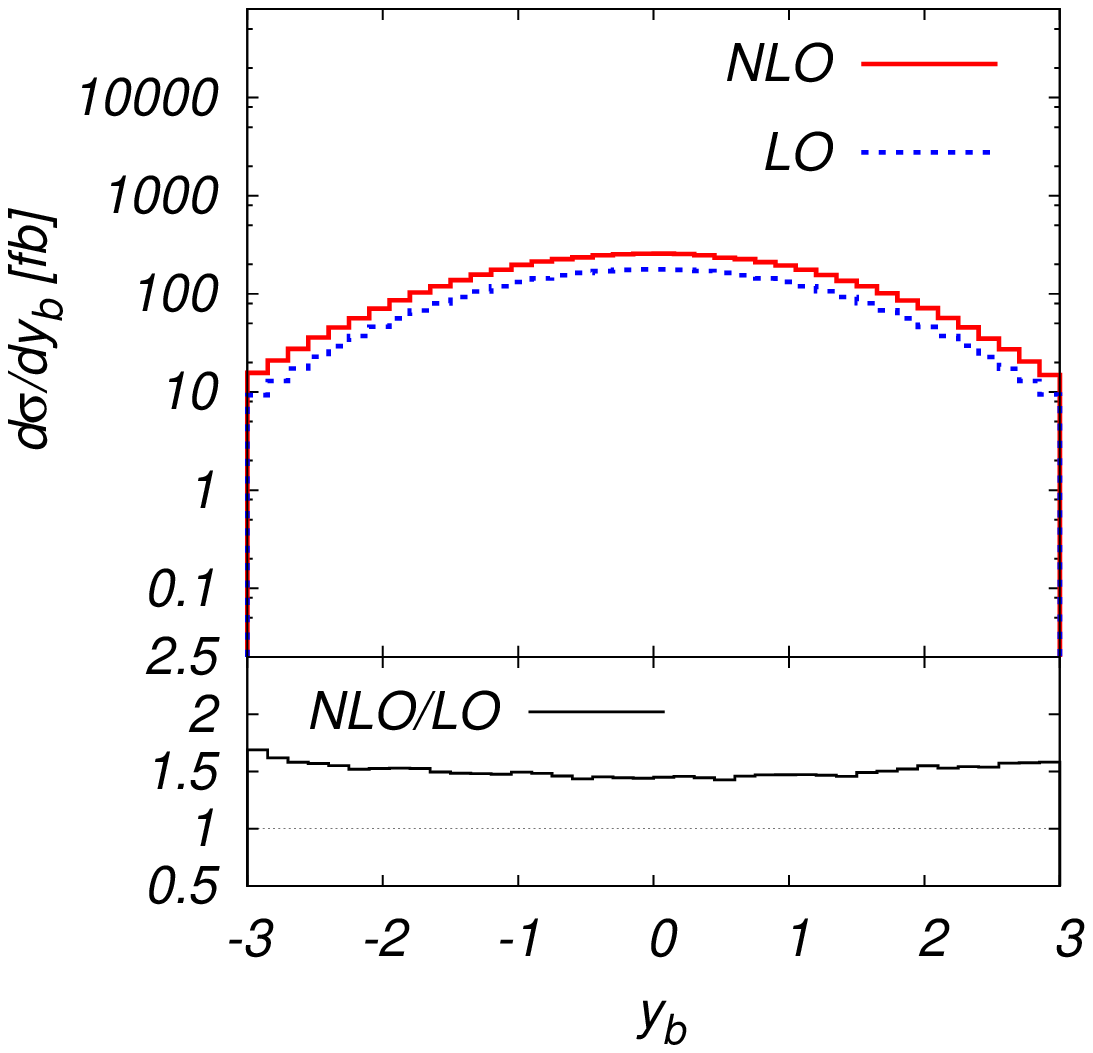}
\includegraphics[width=0.49\textwidth]{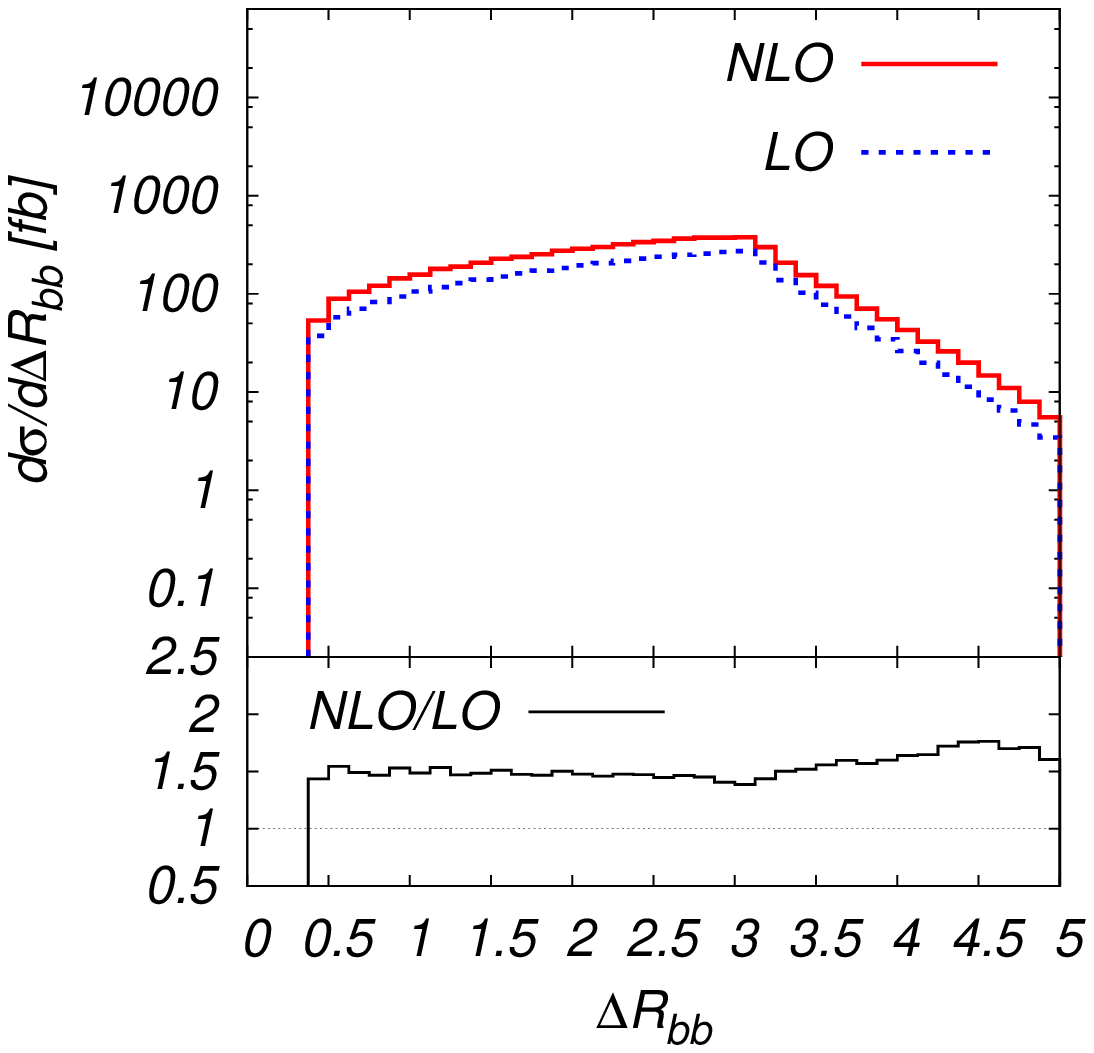}
\end{center}
\caption{\it \label{fig:bottom-lhc} Differential  cross section
  distributions as a function of  the averaged transverse momentum
  $p_{T_{b}}$  of the  b-jet, averaged rapidity  $y_{b}$  of the
  b-jet and $\Delta R_{b\bar{b}}$ separation for the $pp\rightarrow
  e^{+}\nu_{e}\mu^{-}\bar{\nu}_{\mu}b\bar{b} ~ + X$ process at the LHC
  with $\sqrt{s}= 7$ TeV.   The blue dashed curve corresponds to the
  leading order, whereas the red solid one to the next-to-leading
  order result.  The lower panels display  the differential K factor.}
\end{figure}
\begin{figure}[th]
\begin{center}
\includegraphics[width=0.49\textwidth]{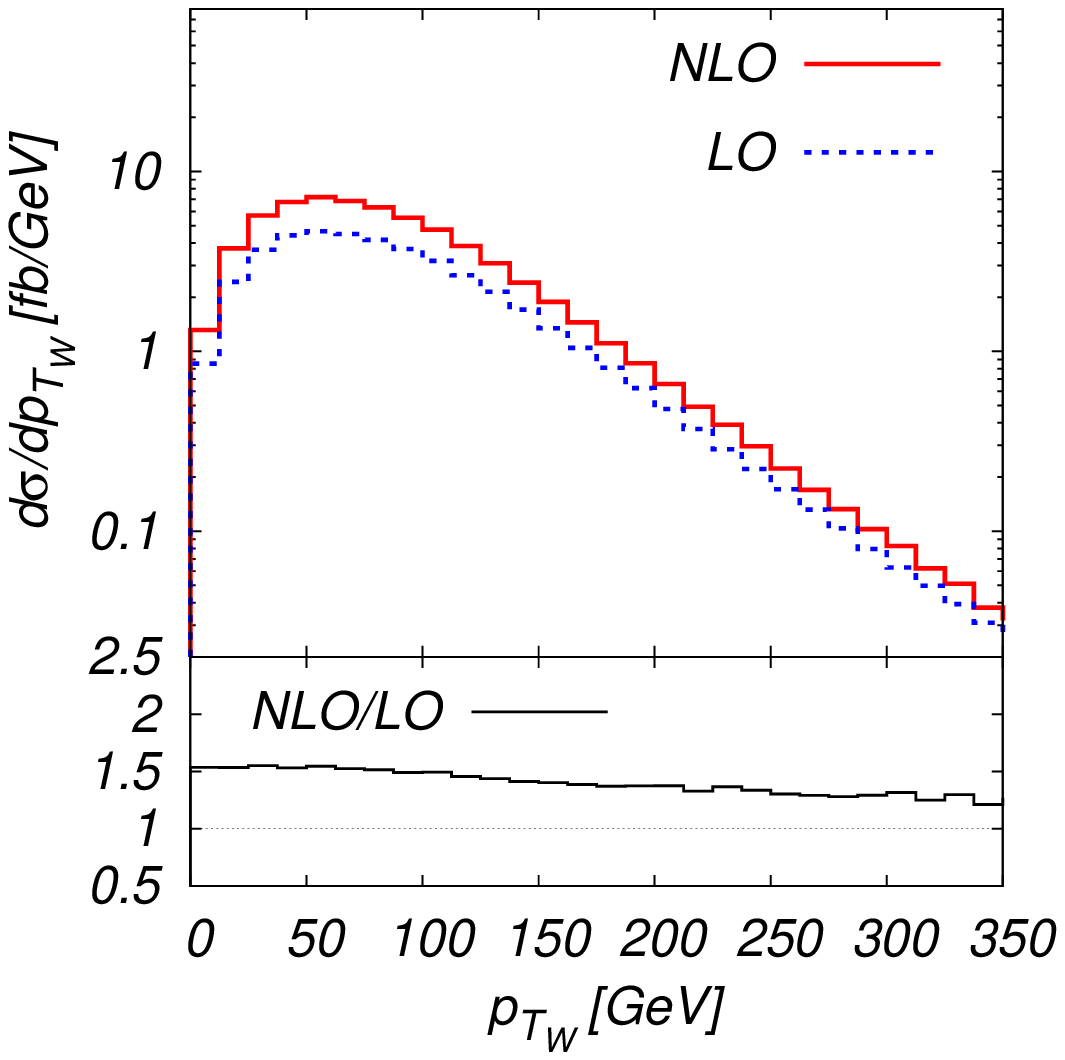}
\includegraphics[width=0.49\textwidth]{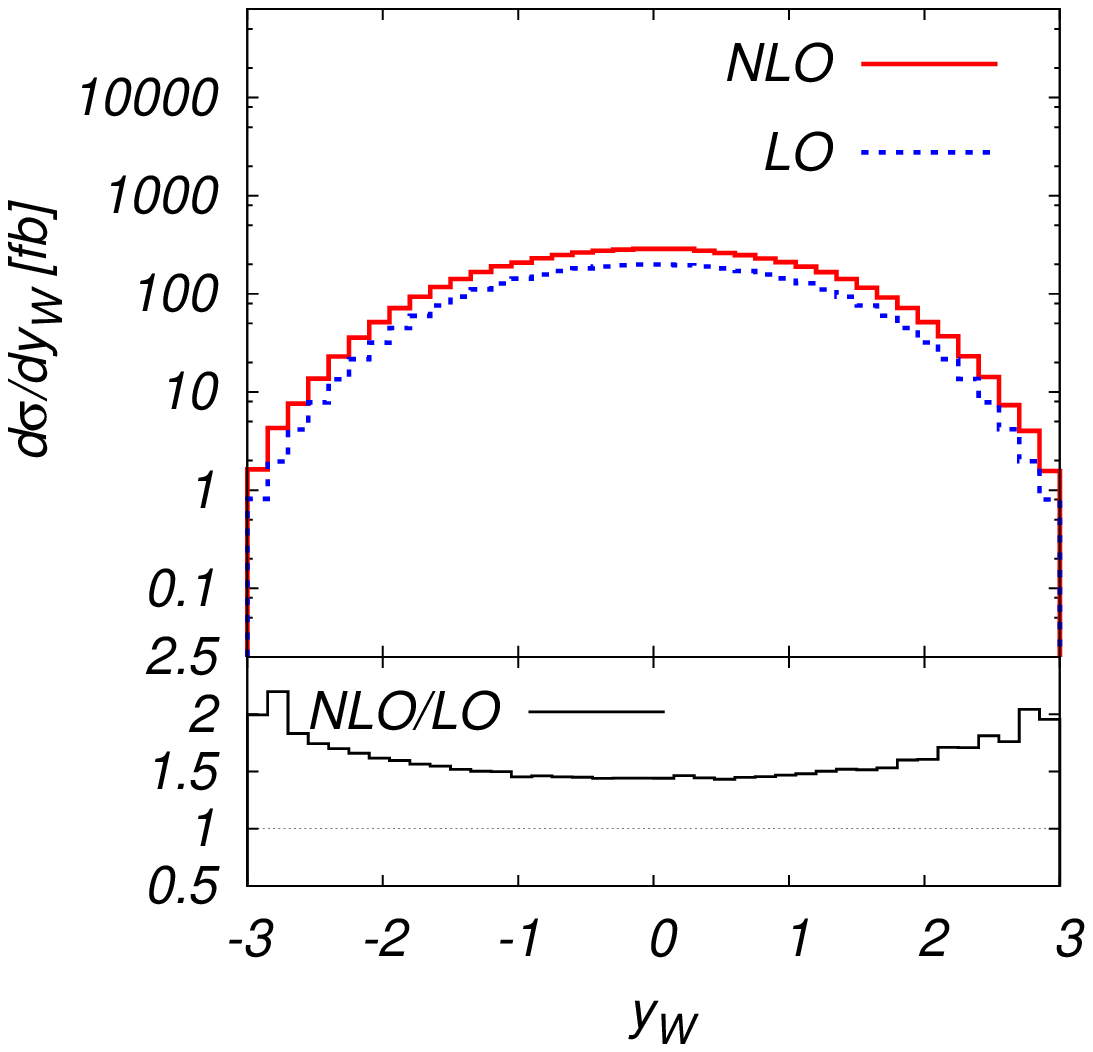}
\end{center}
\caption{\it \label{fig:W-lhc} Differential  cross section
  distributions as a function of the averaged transverse momentum
  $p_{T_{W}}$  of the $W^{\pm}$ bosons and  averaged  rapidity $y_{W}$
  of the $W^{\pm}$ bosons for the $pp\rightarrow
  e^{+}\nu_{e}\mu^{-}\bar{\nu}_{\mu}b\bar{b} ~ + X$ process at the LHC
  with $\sqrt{s}= 7$ TeV.  The blue dashed curve corresponds to the
  leading order, whereas the red solid one to the next-to-leading
  order result. The lower panels display the differential K factor.}
\end{figure}
\begin{figure}[th]
\begin{center}
\includegraphics[width=0.49\textwidth]{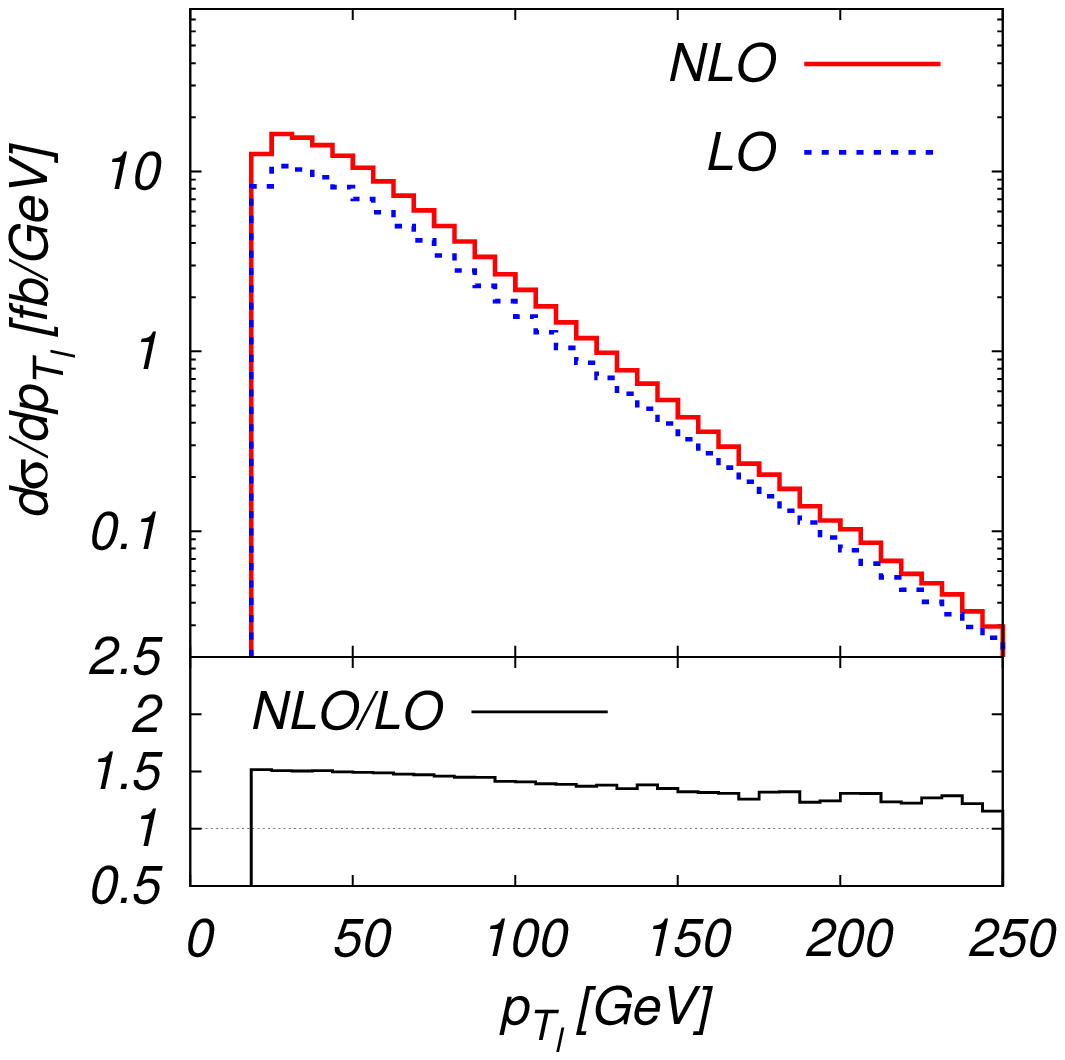}
\includegraphics[width=0.49\textwidth]{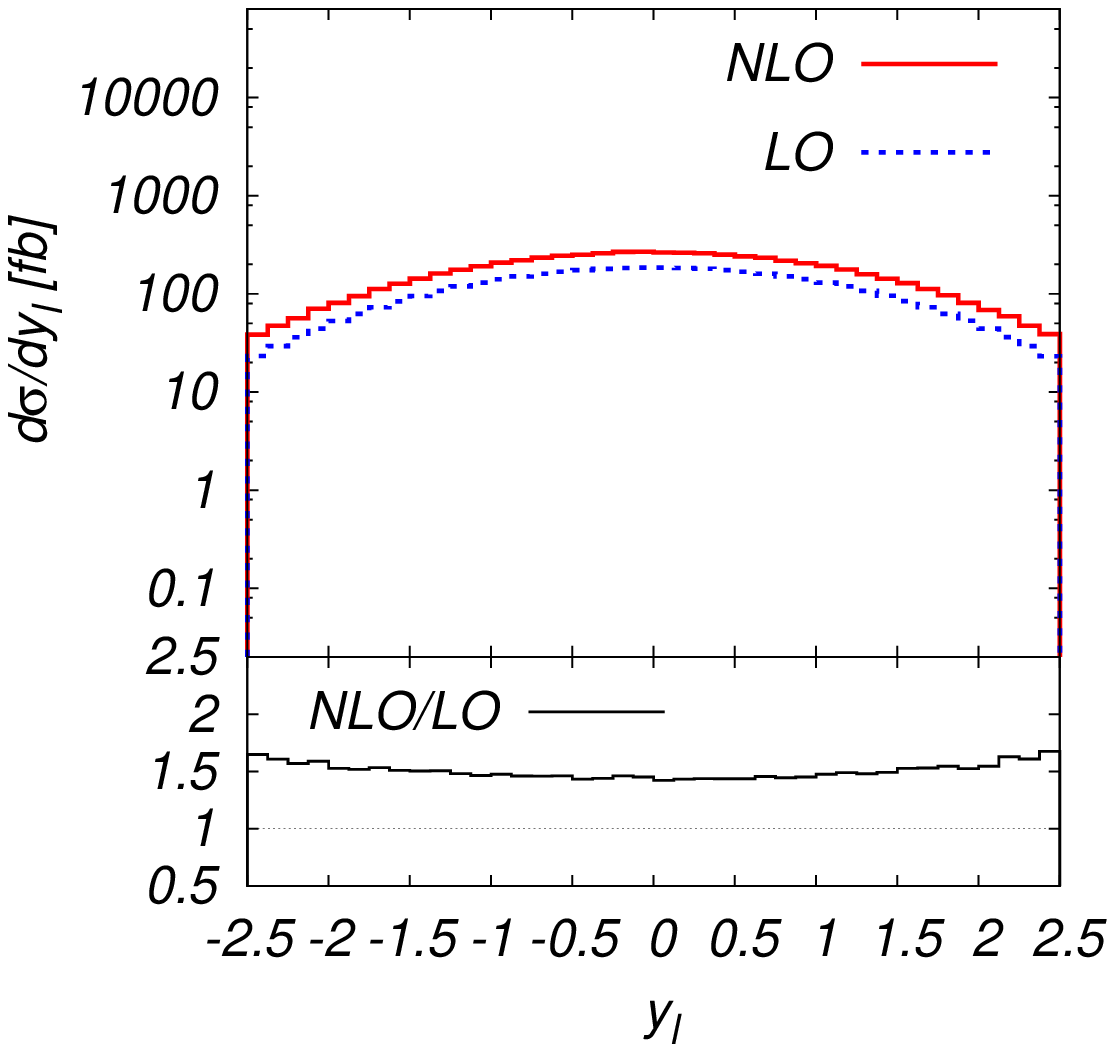}
\includegraphics[width=0.49\textwidth]{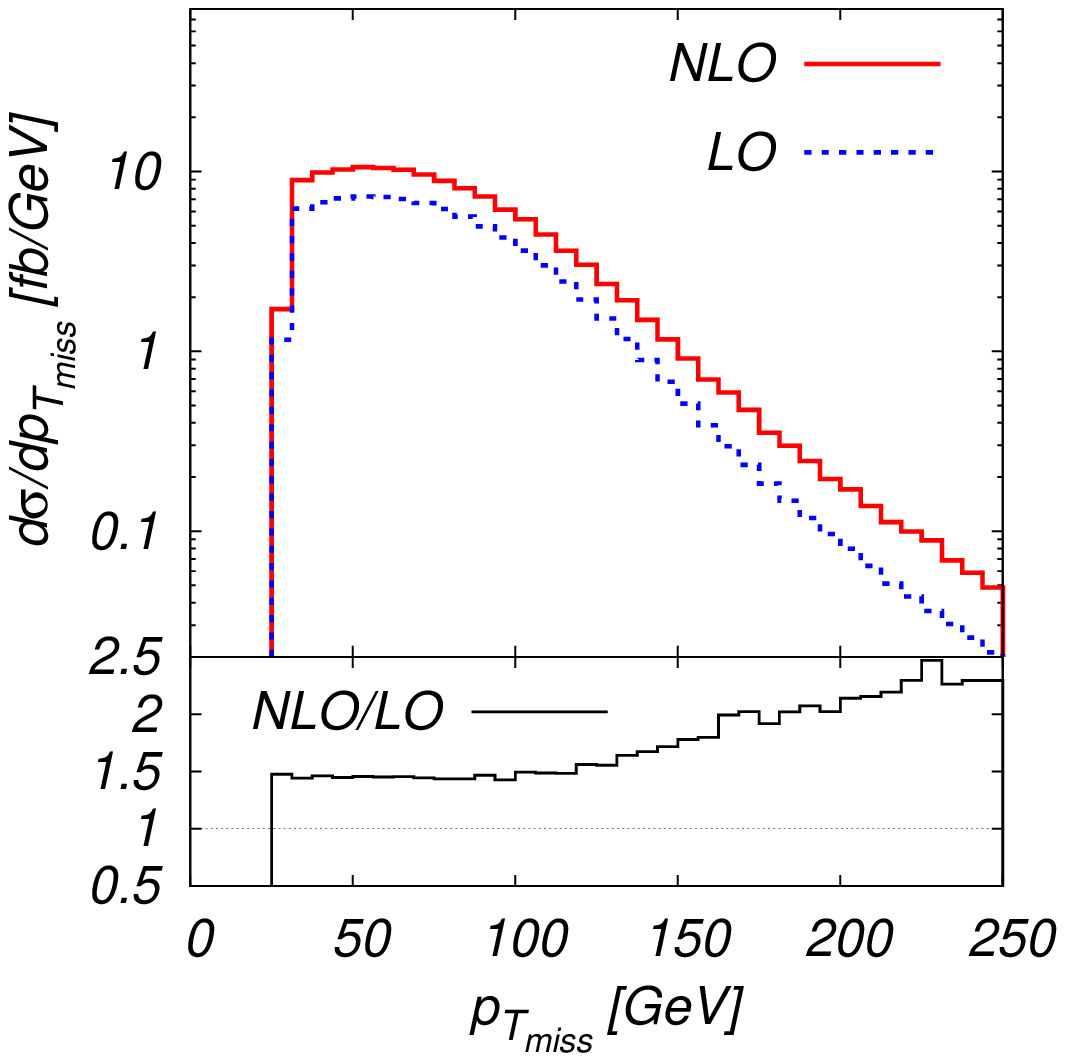}
\includegraphics[width=0.49\textwidth]{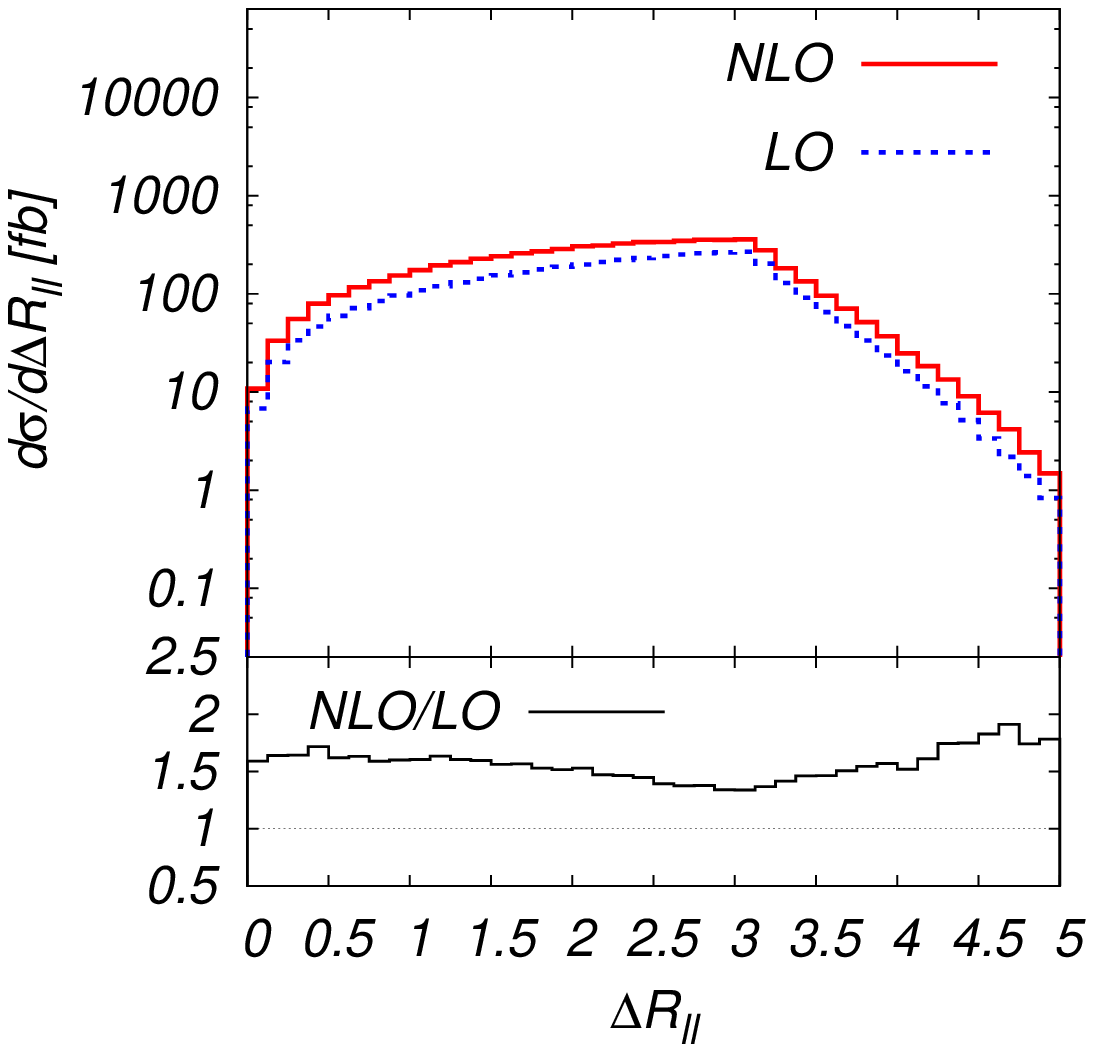}
\end{center}
\caption{\it \label{fig:leptons-lhc} Differential  cross section
  distributions as a function of  the averaged transverse momentum
  $p_{T_{\ell}}$  of the  charged leptons,  averaged  rapidity
  $y_{\ell}$ of the  charged leptons, $p_{T_{miss}}$  and $\Delta
  R_{\ell\ell}$  for the $pp\rightarrow
  e^{+}\nu_{e}\mu^{-}\bar{\nu}_{\mu}b\bar{b} ~ + X$ process at the LHC
  with $\sqrt{s}= 7$ TeV.  The blue dashed curve corresponds to the
  leading order, whereas the red solid one to the next-to-leading
  order result. The lower panels display  the differential K factor. }
\end{figure}
\begin{figure}[th]
\begin{center}
\includegraphics[width=0.49\textwidth]{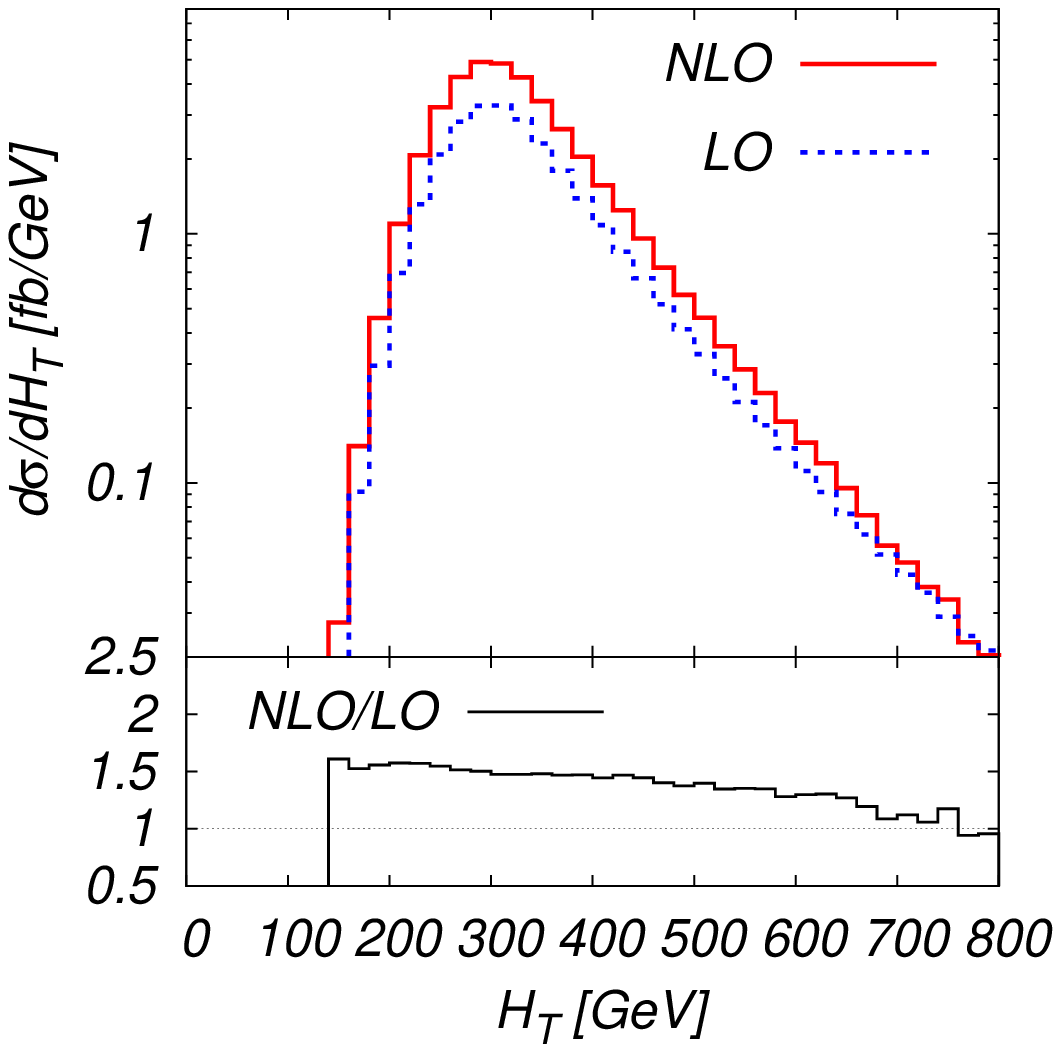}
\end{center}
\caption{\it \label{fig:hT-lhc} Differential  cross section
  distribution as a function of the total transverse energy,  $H_T$,
   for the $pp\rightarrow
  e^{+}\nu_{e}\mu^{-}\bar{\nu}_{\mu}b\bar{b} ~ + X$ process at the LHC
  with $\sqrt{s}= 7$ TeV.  The blue dashed curve corresponds to the
  leading order, whereas the red solid one to the next-to-leading
  order result. The lower panels display the differential K factor.}
\end{figure}

In order to quantify the size of the non-factorizable corrections for
the LHC, we analyze once more the narrow-width limit of our
calculation following the procedure described in Section
\ref{sec:tev}.  Our results are presented in Figure
\ref{fig:rescaling-lhc}, where the dependence of the total NLO cross
section together with its individual contributions, real emission part
and the LO plus virtual corrections, are  shown. Also in this case,
the behavior is compatible with a logarithmic dependence on
$\Gamma_t$ which cancels between real and virtual corrections.  Going
from NWA to the full result changes the cross section no more than
$-1.2\%$ for our inclusive setup, which is  within  the expected
uncertainty of ${\cal{O}}(\Gamma_t/m_t)$ of the NWA approach.

In Table \ref{tab:lhc2}, the integrated cross sections at the LHC with
$\sqrt{s}= 10$ TeV are presented, once more for two choices of the
$\alpha_{max}$ parameter and for the three different jet algorithms.
In this case, at the central scale value, the full cross section
receives   NLO QCD corrections of the order of $43\%$. 

In a next step, we compare our NLO integrated cross section with the
value $\sigma_{\rm NLO} = 2097$ fb for $\sqrt{s}= 10$ TeV
presented in Ref. \cite{Melnikov:2009dn}.  We observe a $5\%$
discrepancy which can perfectly be explained  using the  same
arguments as in the TeVatron case, namely the on-shell top and $W$
boson approximation applied in \cite{Melnikov:2009dn} and small
differences between individual setups. 

As in the case of TeVatron, we have also made a comparison with
\textsc{Mcfm}. We have obtained $\sigma_{\rm LO} = (563.01 \pm 0.63)$
fb, $\sigma_{\rm NLO} = (838.98 \pm 1.68)$ fb for $\sqrt{s} = 7$ TeV
and $\sigma_{\rm LO} = (1421.05 \pm 1.59)$ fb, $\sigma_{\rm NLO} =
(2046.9 \pm 4.3)$ fb for  $\sqrt{s} = 10$ TeV, which constitutes a
difference of $2\%$ at LO for both cases and a difference of $4\%$ and
$3\%$  respectively at NLO. Moreover, both NLO results remain within
our theoretical uncertainty  of $9\%$,  which is due to scale
variation.

Top quark production at the LHC is forward-backward symmetric in the
laboratory frame as a consequence of the symmetric colliding
proton-proton initial state. Therefore, we turn our attention to the
size of NLO QCD corrections to the differential  distributions at the
LHC.

We present the differential distributions only for the $\sqrt{s}= 7$
TeV case.  In Figure \ref{fig:top-lhc}, differential distributions of
the $t\bar{t}$ invariant mass, $m_{t\bar{t}}$, together with  the
rapidity distribution, $y_{t\bar{t}}$,  of the top-anti-top system as
well as the averaged transverse momentum, $p_{T_{t}}$, and  the
averaged  rapidity $y_{t}$ of  the top and anti-top are
depicted. Distributions become harder in $p_T$ and in the invariant
mass of the $t\bar{t}$ pair, moving from the TeVatron to the LHC case,
as expected from the higher scattering energy.  Rapidity distributions
of the  $t\bar{t}$ pair and the $t$ quark,  on the other hand, get
broadened in this transition. NLO QCD corrections  to these
differential distributions are always positive and below
$50\%-60\%$. In case of rapidity distributions this applies for events
concentrated within  $|y_{t\bar{t}}|< 2$  and  $|y_{t}|< 2$  regions.

In Figure \ref{fig:bottom-lhc}, the $b$-jet kinematics is presented
again, but this time in the framework of the LHC. In particular,
differential  cross section  distributions as function of  the
averaged transverse momentum, $p_{T_{b}}$, and averaged rapidity,
$y_{b}$,  of the $b$- and anti-$b$-jet  are presented together with
the $\Delta R_{b\bar{b}}$ separation. Also in this case, the
$p_{T_{b}}$ distribution is harder than at the TeVatron and the
$y_{b}$ distribution is broader. Clearly, the distributions show the
same large and positive corrections, which turn out to be relatively
constant. Only in case of  $\Delta R_{b\bar{b}}$, corrections lead  to
a distortion of the  differential distributions up to  $30\%$.

The $W^{\pm}$ boson kinematics is shown in Figure \ref{fig:W-lhc},
where  the differential  cross section distributions as a function of
the averaged transverse momentum  $p_{T_{W}}$ of the $W^{\pm}$ bosons
together with an averaged  rapidity $y_{W}$ of the $W^{\pm}$ bosons
are depicted. Large positive corrections of $50\%-60\%$ are acquired
for $p_{T_{W}}$ differential distribution and rapidity distribution
with events concentrated within  $|y_{W^{\pm}}|< 2$.  The tails of the
$y_{W^{\pm}}$ distribution acquire even higher NLO QCD corrections.

Subsequently, in Figure \ref{fig:leptons-lhc}, differential  cross
section distributions as function of the averaged transverse momentum
$p_{T_{\ell}}$ and   averaged  rapidity $y_{\ell}$ of the  charged
leptons together with $p_{T_{miss}}$ and $\Delta R_{\ell\ell}$
separation  are shown. A small distortion of the $p_{T_{\ell}}$
differential distribution up to $25\%$ is reached, while for
$p_{T_{miss}}$ a distortion up to  $70\%-80\%$ is visible.  For
the $y_{\ell}$ distribution, large positive and rather constant
corrections up to $50\%$  are obtained, and for the tails of the
$\Delta R_{\ell\ell}$ distribution corrections of $80\%-90\%$ are
obtained. 

Finally, in Figure \ref{fig:hT-lhc} the differential   cross section
distribution as function of the total transverse energy defined in
(\ref{defHT}) is presented. In this case we observe a distortion of
the differential distribution up to  $40\%$.

Generally, we can say that for a fixed scale $\mu=m_t$ at LHC, the NLO
QCD corrections are  always positive and large, at the level of
$50\%-60\%$. Furthermore, they are  relatively constant. Exceptions
are  the rapidity distributions, which are only constant in the
central region, and the $p_{T_{miss}}$  and $H_T$ distributions,
which are distorted up to $40\%-80\%$.

\section{Conclusions}
\label{sec:conclusions}

In this paper, we have presented, for the first time, a computation of
the NLO QCD corrections to the full  decay  chain $pp(p\bar{p})
\rightarrow t \bar{t}\to W^+W^- b\bar{b} \to e^{+}\nu_{e}
\mu^{-}\bar{\nu}_{\mu}b\bar{b} ~+X$. All off-shell effects of top
quarks and $W$  gauge  bosons have been included in a fully
differential manner which  allows us to compute an arbitrary
observable in terms of jets, charged leptons and missing transverse
energy within experimentally relevant selection criteria with NLO QCD
accuracy.  In order to illustrate the capabilities of the program, the
total cross section and its scale dependence, as well as several
differential distributions at the TeVatron run II and the LHC have
been given. Moreover, in case of the TeVatron the forward-backward
asymmetry of the top has been recalculated.  We have found that with
inclusive selection cuts,  the forward-backward asymmetry  amounts to
$A^{t}_{FB} = 0.051 \pm 0.0013$. Furthermore, the  impact of the NLO
QCD corrections  on integrated cross sections at the TeVatron  is
small,  of the order $2.3 \%$. At the LHC we have obtained NLO QCD
corrections at the level of $47\%$ and $43\%$  for $\sqrt{s}= 7$ TeV
and $\sqrt{s}= 10$ TeV respectively. A study of the scale dependence
of our NLO predictions indicates that the residual theoretical
uncertainty due to higher order corrections is $8\%$ for the TeVatron
and $9\%$ for the LHC.  

\section*{Acknowledgments}

We would like to thank John Campbell and Keith Ellis  for their help
in using \textsc{Mcfm}.

The work was funded in part by the RTN European Programme
MRTN-CT-2006-035505 HEPTOOLS - Tools and Precision Calculations for
Physics Discoveries at Colliders. 

M. Czakon was supported by the Heisenberg and by the Gottfried
Wilhelm Leibniz Programmes of the Deutsche Forschungsgemeinschaft and
M. Worek by the Initiative and Networking Fund of the Helmholtz
Association, contract HA-101 ("Physics at the Terascale").  

\section*{Note added}
Independently of our calculation, another group has evaluated NLO QCD
corrections to WWbb production with leptonic decays of gauge bosons,
and has presented them in \cite{Denner:2010jp}. We have cross checked
the results, applying the narrow width approximation for the W bosons
as in that publication, and have obtained perfect agreement for
integrated LO and NLO cross sections within statistical errors.

\providecommand{\href}[2]{#2}
\begingroup\raggedright\endgroup

\end{document}